\documentclass[preprint,11pt,floatfix,showpacs,showkeys,preprintnumbers,amsmath,amssymb,nofootinbib]{revtex4}

\usepackage{graphicx}
\usepackage{dcolumn}
\usepackage{bm}
\usepackage{longtable}
\usepackage{latexsym}
\newcommand{\ket}[1]{|{#1}\rangle}
\newcommand{\bra}[1]{\langle {#1}|}
\newcommand{\vectorjc}[2]{j\sp{{#1}}\sb{V{#2}}}
\newcommand{\axialjc}[2]{j\sp{{#1}}\sb{A{#2}}}
\newcommand{\mpi}{m\sb{\pi}}
\newcommand{\fpi}{f\sb{\pi}}
\newcommand{\isomtx}[3]{{I}\sp{#1}({\textstyle \frac{#2}{2},\frac{#3}{2}})}

\begin{document}

\title{
The $\pi N\rightarrow \pi\pi N$ reaction around the $N\sp{\ast}(1440)$ energy
}

\author{Hiroyuki Kamano}
\email{kamano@kern.phys.sci.osaka-u.ac.jp}
\affiliation{Department of Physics, Osaka University, Toyonaka, Osaka 560-0043, Japan}
\author{Masaki Arima}
\email{arima@ocunp.hep.osaka-cu.ac.jp}
\affiliation{Department of Physics, Osaka City University, Osaka 558-8585, Japan}

\date{\today}

\begin{abstract}
We study the $\pi N \rightarrow \pi \pi N$ reaction
around the $N\sp{\ast}(1440)$ mass-shell energy.
Considering the total cross sections and invariant
mass distributions, we discuss the role
of $N\sp{\ast}(1440)$ and its decay processes on this reaction.
The calculation is performed by extending our previous approach
[Phys. Rev. C \textbf{69}, 025206 (2004)],
in which only the nucleon and $\Delta(1232)$ were
considered as intermediate baryon states.
The characteristics in the recent data
of the $\pi\sp{-} p \rightarrow \pi\sp{0} \pi\sp{0} n$ reaction
measured by Crystal Ball Collaboration (CBC),
can be understood as a strong interference between the two decay processes: 
$N\sp{\ast}(1440) \rightarrow \pi \Delta$ and
$N\sp{\ast}(1440) \rightarrow N(\pi \pi)\sp{I=0}\sb{S\text{ wave}}$.
It is also found that the scalar-isoscalar $\pi \pi$ rescattering effect 
in the $NN\sp{\ast}(\pi \pi)\sp{I=0}\sb{S\text{ wave}}$ vertex, 
which corresponds to the propagation of $\sigma$ meson,
seems to be necessary for explaining the several observables of 
the $\pi N \rightarrow \pi \pi N$ reaction: 
the large asymmetric shape in the $\pi\sp{0}\pi\sp{0}$ invariant mass 
distributions of the $\pi\sp{-} p \rightarrow \pi\sp{0}\pi\sp{0}n$ reaction 
and the $\pi\sp{+} p \rightarrow \pi\sp{+} \pi\sp{+} n$ total cross section.
\end{abstract}



\maketitle
%
%
%
%
%
%
%
%
\section{Introduction}
\label{sec1}
In many pion induced reactions on the nucleon,
the single pion production reaction $\pi N \rightarrow \pi \pi N$ 
has been studied with a particular interest 
due to its role as a major inelastic process. 
The partial wave analyses of this reaction, 
together with other reaction channels such as 
the elastic and $\pi N \rightarrow \eta N$ reactions,
 have revealed various properties
of the nonstrange baryon resonances, $N\sp{\ast}$'s and $\Delta$'s 
(e.g. see Refs.~\cite{Man84,Man92,Arn04}).
A number of theoretical investigations have also been performed
on the basis of the phenomenological approaches
using the effective Lagrangian~\cite{Jen97,Jak90,Jak92,Jak93,Ose85}
and the chiral perturbation theory~\cite{Ber95,Ber97,Fet00,Mob05}.

In Ref.~\cite{Kam04}, we have discussed the total cross sections of 
this reaction up to $T\sb{\pi}=400$ MeV, making use of 
the chiral reduction formula proposed by Yamagishi and Zahed~\cite{Yam96}. 
We focused on the role of $\Delta(1232)$,
i.e. the influence of $\pi \Delta \Delta$ and
$\rho N \Delta$ interactions on the reaction processes. 
Because these interactions are not directly observed through the two-body decay
of $\Delta(1232)$ in contrast to the $\pi N \Delta$ interaction,
their coupling constants are difficult to determine.
We found that the $\pi\sp{\pm} p \rightarrow \pi\sp{\pm} \pi\sp{0} p$
reactions are sensitive to the $\pi \Delta \Delta$ and
$\rho N \Delta$ interactions, 
and could be a source of information of their coupling constants.

Besides being useful for clarifying the properties of $\Delta(1232)$,
the $\pi N \rightarrow \pi \pi N$ reaction is expected 
to provide us valuable information about the Roper resonance $N\sp{\ast}(1440)$
and its decay to the $\pi \pi N$ channel.
$N\sp{\ast}(1440)$ decays to the $\pi \pi N$ channel
via the $N\sp{\ast}(1440) \rightarrow \pi \Delta$
and $N\sp{\ast}(1440) \rightarrow N(\pi\pi)\sp{I=0}\sb{S\text{ wave}}$
processes,
which have the branching ratios of about 25~\% and 7.5~\%, 
respectively~\cite{PDG04}.
The importance of the latter process 
has already been pointed out in several studies~\cite{Jen97,Ber95}.

The importance of $N\sp{\ast}(1440)$ and its subsequent decays
has been discussed extensively also in 
the $pp\rightarrow pp\pi\sp{+}\pi\sp{-}$ and
$pn\rightarrow d(\pi\pi)\sp{I=0}$ 
reactions~\cite{Alv98,Alv99,Alv01,Bro02,Pat03}.
However,
such two-pion decay of $N\sp{\ast}(1440)$ is not easy to study
in the elastic $\pi N \rightarrow \pi N$ scattering,
and also in the $\gamma N \rightarrow \pi \pi N$ reaction
because $N\sp{\ast}(1440)$ has small electromagnetic transition
rate compared to other relevant resonances
such as $\Delta(1232)$ and $N\sp{\ast}(1520)$~\cite{Gom96}.

Recently, the $\pi\sp{-} p \rightarrow \pi\sp{0}\pi\sp{0}n$ reaction 
has been measured up to $p\sb{\pi\sp{-}}=750$~MeV/c 
(i.e. $T\sb{\pi}\sim 620$~MeV)
by Crystal Ball Collaboration (CBC)~\cite{Pra04,Cra03}.
These high precision data cover the energy region far from the threshold,
in particular around the $N\sp{\ast}(1440)$ mass-shell energy,
$T\sb{\pi}\sim 480$~MeV. 
Several new interesting results related to $N\sp{\ast}(1440)$
and its decay processes were reported:
(i)
The energy dependence of the total cross section shows 
a shoulder-like shape just below the $N\sp{\ast}(1440)$ energy 
(see Fig.~14 in Ref.~\cite{Pra04}, and it should be compared with
the $\gamma p \rightarrow \pi\sp{0} \pi\sp{0} n$ total cross section
displayed in the same figure, for which no such shape appears because
of the small radiative coupling of $N\sp{\ast}(1440)$). 
(ii)
The $\pi\sp{0} n$ invariant mass distribution shows a peak
near the invariant mass equal to the $\Delta(1232)$ energy,
which would be produced via
the process $N\sp{\ast}(1440) \rightarrow \pi\Delta$.
(iii)
The $\pi\sp{0}\pi\sp{0}$ invariant mass distributions
shows a large asymmetric shape in population of the events, i.e.
the peak at large value of $m\sp{2}(\pi\sp{0} \pi\sp{0})$ is larger than
that at small value of $m\sp{2}(\pi\sp{0} \pi\sp{0})$.

When the total energy increases,
the correlation between the outgoing pions
will become visible.
The scalar-isoscalar correlation of two pions via 
the $N\sp{\ast}(1440) \rightarrow N(\pi\pi)\sp{I=0}\sb{S\text{ wave}}$ decay 
is particularly interesting,
because such correlation may generate the $\sigma$ meson pole.
In view of the sizable contribution of 
$N\sp{\ast}(1440) \rightarrow N(\pi\pi)\sp{I=0}\sb{S\text{ wave}}$,
the $\pi N \rightarrow \pi \pi N$ reaction 
is a possible source of information 
about this controversial scalar meson.
It is worth noting that several literatures have suggested that 
this ``$\sigma$'' degree of freedom is important 
also in understanding the structure of $N\sp{\ast}(1440)$~\cite{Kre00,Dil04}. 

In this paper, we investigate the $\pi N \rightarrow \pi \pi N$ reaction 
in the energy region up to $T\sb{\pi} = 620$ MeV, especially around the
$N\sp{\ast}(1440)$ mass-shell energy.
Through the comparison with the recent CBC data, 
we particularly discuss the role of $N\sp{\ast}(1440)$ 
and its decay processes on this reaction.
Furthermore, we try to discuss the possibility of extracting 
the information about the $\sigma$ meson such as its existence.
The calculation is performed by extending and improving
the theoretical framework of our previous study~\cite{Kam04}
in which the contributions of $\Delta(1232)$ 
have been discussed in detail in the energy region up to $T\sb{\pi}=400$~MeV.

This paper is organized as follows.
In Sec.~\ref{sec2}, we give a brief summary of our previous study
and explain new ingredients introduced in this work.
The numerical results are presented in Sec.~\ref{sec3} and 
the contributions of $N\sp{\ast}(1440)$ to 
the $\pi N \rightarrow \pi \pi N$ reaction are discussed.
Then our results are compared with the recent CBC data.
Summary and conclusions are given in Sec.~\ref{sec4}.
In the Appendices we summarize some details of phenomenological treatment 
in the calculation.
\section{Theoretical Treatment of $\pi N \rightarrow \pi \pi N$ reaction}
\label{sec2} 
\subsection{Brief summary of our previous study}
\label{sec2-1}
The starting point of our previous study~\cite{Kam04}
is the chiral Ward identity satisfied by 
the invariant amplitude ${\cal M}\sb{\pi\pi N}$.
\begin{figure}
\includegraphics[width = 5cm]{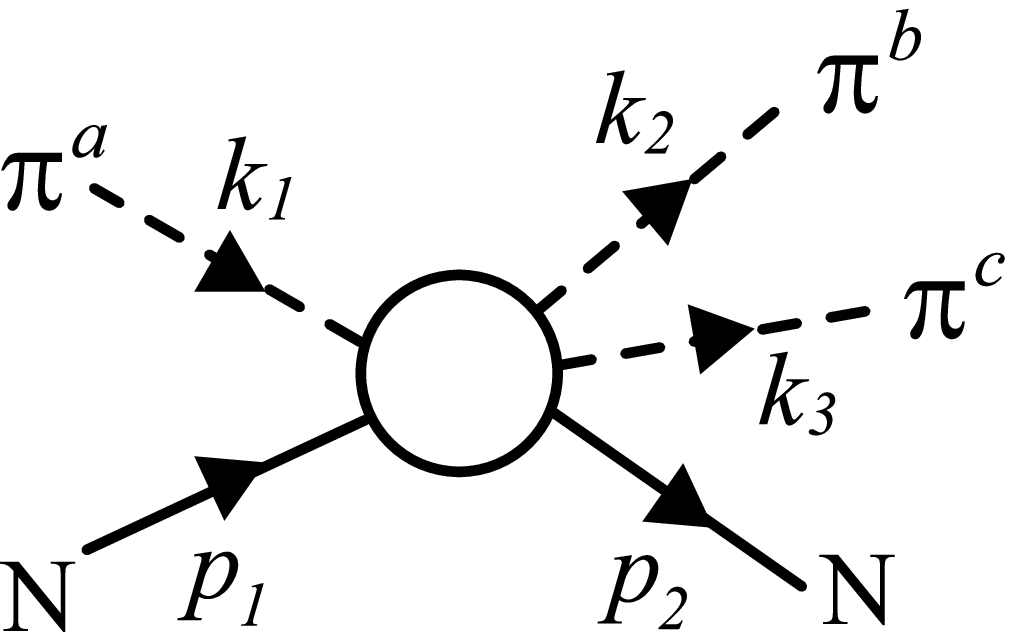}
\caption{
The $\pi N \rightarrow \pi \pi N$ reaction.
Pions have the isospin index ($a,b,c$) and the four-momentum 
$k\sb{i}$ ($i=1,2,3$), and nucleons have the four-momentum
$p\sb{j}$ ($j=1,2$).
}
\label{fig1}
\end{figure}
Assigning the four-momentum and isospin indices
of the external nucleons and pions as in Fig.~\ref{fig1},
we have
\begin{eqnarray}
{\cal M}\sb{\pi\pi N} &=&
( {\cal M}\sb{\pi}+ {\cal M}\sb{A}
+ {\cal M}\sb{SA} + {\cal M}\sb{VA} )
\nonumber\\ 
&&
+ (k\sb{1},a \leftrightarrow -k\sb{3},c)
+ (k\sb{2},b \leftrightarrow  k\sb{3},c)
\nonumber\\
&&
+ {\cal M}\sb{AAA},
\label{eq1}
\end{eqnarray}
where the symbol $(\ \leftrightarrow\ )$ represents a permutation of 
the momentum and isospin indices of the pion in the first four terms,
and
\begin{equation}
{\cal M}\sb{\pi} =
\frac{1}{\fpi\sp{2}} [(k\sb{1}-k\sb{2})\sp{2} - m\sb{\pi}\sp{2}]
\delta\sp{ab}
\bra{N(p\sb{2})}\hat\pi\sp{c}(0)\ket{N(p\sb{1})},
\label{eq2}
\end{equation}
\begin{equation}
{\cal M}\sb{A} =
-\frac{i}{2f\sb{\pi}\sp{3}}(k\sb{2}-k\sb{1})\sp{\mu}\delta\sp{ab}
\bra{N(p_2)}\axialjc{c}{\mu}(0)\ket{N(p_1)},
\label{eq3}
\end{equation}
\begin{eqnarray}
{\cal M}\sb{SA} &=&
- \frac{\mpi\sp{2}}{\fpi\sp{2}}k\sb{3}\sp{\mu} \delta\sp{ab}
\int d\sp{4}x e\sp{-i(k\sb{1} - k\sb{2}) x}
\nonumber\\
&&
\times
\bra{N(p\sb{2})}
T\sp{\ast} \bm{(} \hat{\sigma}(x)\axialjc{c}{\mu}(0) \bm{)}
\ket{N(p\sb{1})},
\label{eq4}
\end{eqnarray}
\begin{eqnarray}
{\cal M}\sb{VA} &=&
-\frac{i}{2\fpi\sp{3}} (k\sb{1} + k\sb{2})\sp{\mu} k\sb{3}\sp{\nu}
\varepsilon\sp{abe}
\int d\sp{4}x e\sp{-i(k\sb{1} - k\sb{2}) x}
\nonumber\\
&&
\times
\bra{N(p\sb{2})}
T\sp{\ast} \bm{(} \vectorjc{e}{\mu}(x)\axialjc{c}{\nu}(0) \bm{)}
\ket{N(p\sb{1})},
\label{eq5}
\end{eqnarray}
\begin{eqnarray}
{\cal M}\sb{AAA} &=&
\frac{i}{f\sb{\pi}\sp{3}}k\sb{1}\sp{\mu}k\sb{2}\sp{\nu}k\sb{3}\sp{\lambda}
\int d\sp{4}x\sb{1} d\sp{4}x\sb{2} e\sp{-ik\sb{1}x\sb{1} + ik\sb{2}x\sb{2}}
\nonumber\\
&&
\times
\bra{N(p_2)}T\sp{\ast} 
\bm{(}\axialjc{a}{\mu}(x_1)\axialjc{b}{\nu}(x_2)\axialjc{c}{\lambda}(0)\bm{)}
\ket{N(p_1)}.
\label{eq6}
\end{eqnarray}
The pseudoscalar density $\hat{\pi}\sp{a}(x)$ 
is the interpolating pion field with the asymptotic form 
$\hat{\pi}\sp{a}(x) \rightarrow \pi\sb{\text{in,out}}(x) + \cdots$ 
$(x\sb{0}\rightarrow \mp \infty)$. 
The one-pion reduced axial current $\axialjc{a}{\mu}(x)$ is 
defined by 
$\axialjc{a}{\mu}(x)=A\sp{a}\sb{\mu}(x)
+f\sb{\pi}\partial\sb{\mu}\hat{\pi}\sp{a}(x)$ 
where $A\sp{a}\sb{\mu}(x)$ 
is the ordinary axial current with the asymptotic form
$A\sp{a}\sb{\mu}(x)\rightarrow 
-f\sb{\pi}\partial\sb{\mu}\pi\sb{\text{in,out}}(x)+\cdots$
$(x\sb{0}\rightarrow \mp \infty)$.
The vector current and the scalar density are represented by
$\vectorjc{a}{\mu}(x)$ and $\hat{\sigma}(x)$, respectively.

The Ward identity~(\ref{eq1})-(\ref{eq6}) was first derived 
by Yamagishi and Zahed making use of 
the chiral reduction formula~\cite{Yam96,Ste98}.
Owing to this formula the invariant amplitude is expressed in terms of 
Green's functions of well-defined current and density operators.
Then the consequences of broken chiral symmetry subject to the 
asymptotic condition  
$\partial\sp{\mu} A\sp{a}\sb{\mu}(x)\rightarrow 
f\sb{\pi}\mpi\sp{2}\pi\sb{\text{in,out}}(x)+\cdots$
$(x\sb{0}\rightarrow \mp \infty)$, are exactly embodied
on the amplitude without relying on any specific model or expansion scheme.
Thus, by using the chiral reduction formula at the beginning of the discussion,
we can consider the detail of each reaction mechanism separately from the 
general framework required by broken chiral symmetry. 
This nature of the chiral reduction formula has a great advantage
in tackling on the hadronic processes in the resonance region
in which the systematic chiral expansion scheme becomes difficult
to be implemented.

\begin{figure}
\includegraphics[width=8cm]{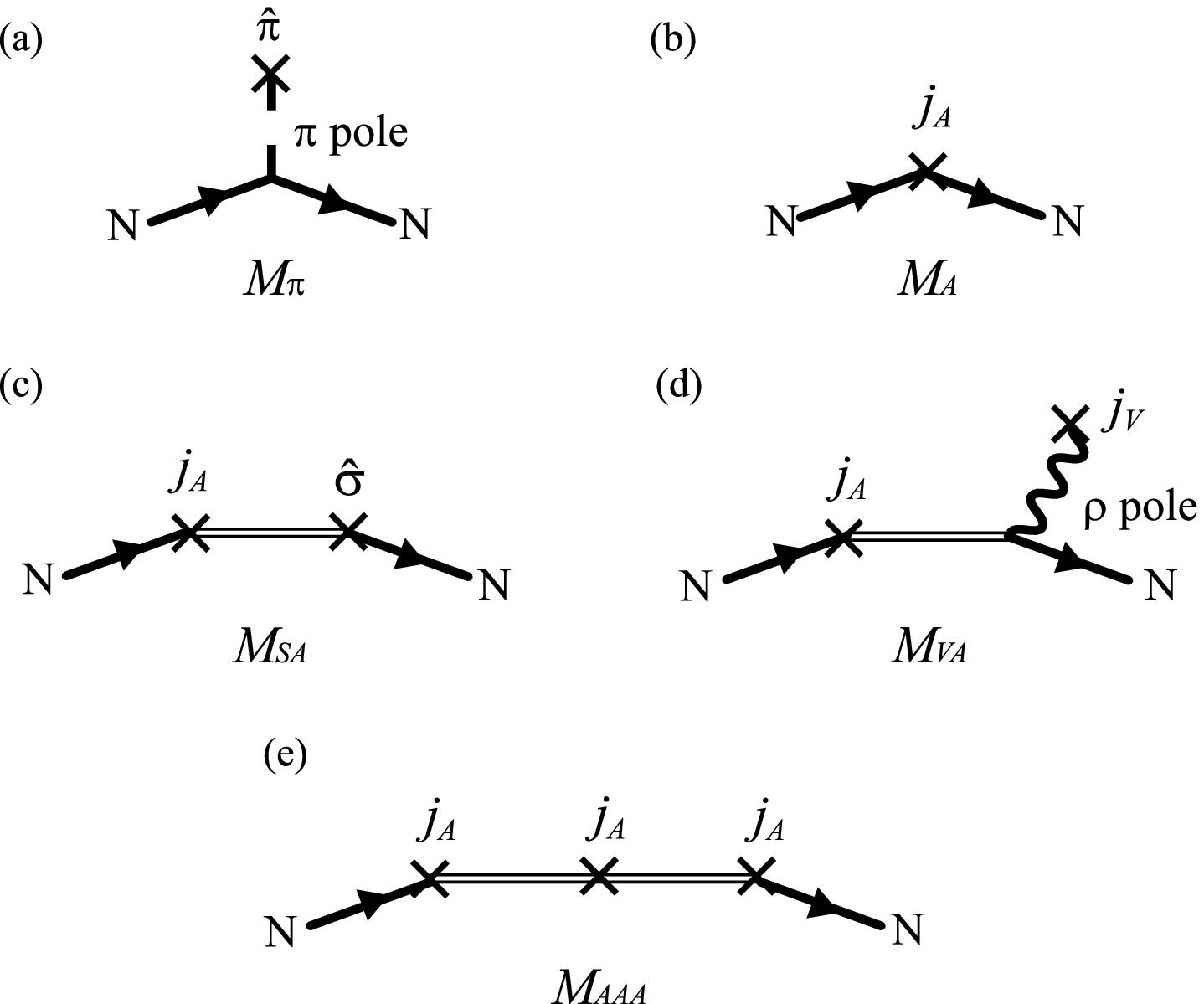}
\caption{
The diagrams considered in our previous study~\cite{Kam04}
to evaluate Green's functions in Eqs.~(\ref{eq2})-(\ref{eq6}).
Crossed versions are also considered for all of these diagrams.
The double line represents the propagation of the nucleon or $\Delta(1232)$.
Note that $\Delta(1232)$ does not propagate in ${\cal M}\sb{SA}$ because the
$N$-$\Delta$ transition is not brought about by 
the scalar density operator $\hat{\sigma}$.
The pseudoscalar density $\hat{\pi}$ and the vector current $\vectorjc{}{}$
are dominated by the pion and $\rho$ meson pole, respectively.
}
\label{fig2}
\end{figure}

Green's functions (i.e. the matrix elements of current and density operators)
appearing in Eqs.~(\ref{eq2})-(\ref{eq6})
are not uniquely determined by broken chiral symmetry.
Therefore we need to employ a model in order to evaluate Green's functions.
In Ref.~\cite{Kam04} they were calculated by taking a phenomenological 
approach based on the relativistic tree level diagrams
as shown in Fig.~\ref{fig2}.
We considered only the nucleon and $\Delta(1232)$ 
as the intermediate baryons, and $\pi$ and $\rho$ as the
intermediate mesons.
Details of our model~\cite{Kam04} are summarized in Appendix~\ref{app1}.
\subsection{Contribution of $N\sp{\ast}(1440)$}
\label{sec2-2}
Now we extend our previous approach by including $N\sp{\ast}(1440)$. 
As for the contributions of $N\sp{\ast}(1440)$ on the reaction,
we consider the $\pi N N\sp{\ast}$, 
$N N\sp{\ast} (\pi\pi)\sp{I=0}\sb{S\text{ wave}}$,
and $\pi \Delta N\sp{\ast}$ interactions.
The first two interactions have already been considered 
in many theoretical investigations of the $\pi N \rightarrow \pi \pi N$
reaction near threshold.
While the $\pi \Delta N\sp{\ast}$ interaction has often been neglected,
this interaction will become important
in the energy region considered in this paper.
Indeed, the importance of $N\sp{\ast} \rightarrow \Delta \pi$ decay
is suggested experimentally~\cite{Cra03,Pra04}.
\subsubsection{
$\pi N N\sp{\ast}$ and $\pi \Delta N\sp{\ast}$ interactions}
\label{sec2-2-1}
The $\pi N N\sp{\ast}$ and $\pi \Delta N\sp{\ast}$ vertices 
with the one-pion leg are generally related to 
the matrix elements of axial current $\axialjc{a}{\mu}$
for the $N$-$N\sp{\ast}$ transition as
\begin{eqnarray}
&&
\bra{N(p\sp{\prime})} \axialjc{a}{\mu}(0) \ket{N\sp{\ast}(p)}
\nonumber\\
&=&
\bar{u}\sb{N}(p\sp{\prime})
\left[ 
  F\sp{N N\sp{\ast}}\sb{A,1}(t)\gamma\sb{\mu} 
+ F\sp{N N\sp{\ast}}\sb{A,2}(t)q\sb{\mu}
\right]
\gamma\sb{5} \frac{\tau\sp{a}}{2}
u\sb{N\sp{\ast}}(p),
\nonumber\\
&&
\label{eq7}
\end{eqnarray}
and for the $\Delta$-$N\sp{\ast}$ transition as
\begin{eqnarray}
&&
\bra{\Delta(p\sp{\prime})}\axialjc{a}{\mu}(0)\ket{N\sp{\ast}(p)} 
\nonumber\\
&=&
\bar{U}\sp{\nu}(p\sp{\prime})
\left[ 
  F\sp{\Delta N\sp{\ast}}\sb{A,1}(t) g\sb{\nu\mu}
+ F\sp{\Delta N\sp{\ast}}\sb{A,2}(t) Q\sb{\nu} \gamma\sb{\mu}
\right.
\nonumber\\
&&
\left.
+ F\sp{\Delta N\sp{\ast}}\sb{A,3}(t) Q\sb{\nu} Q\sb{\mu}
+ F\sp{\Delta N\sp{\ast}}\sb{A,4}(t) 
  Q\sb{\nu} i\sigma\sb{\mu\lambda} Q\sp{\lambda}
\right]
I\sp{a}({\textstyle \frac{3}{2},\frac{1}{2}})
u\sb{N\sp{\ast}}(p),
\label{eq8}
\end{eqnarray}
respectively, 
where $q\sp{\mu}=(p\sp{\prime}-p)\sp{\mu}$, $Q\sp{\mu}=-q\sp{\mu}$
and $t=(p\sp{\prime}-p)\sp{2}$,
$\tau\sp{a}$ is the isospin Pauli matrix and $I\sp{a}(i,j)$ is 
the $j\rightarrow i$ isospin transition $(2i+1)\times(2j+1)$ matrix.
The isoquadruplet Rarita-Schwinger vector-spinor 
and the isodoublet Dirac spinor are denoted as
$U\sp{\mu}(p)$ and $u\sb{i}(p)$ ($i=N,N\sp{\ast}$), respectively. 
By employing Eqs.~(\ref{eqb6}) and (\ref{eqb7}) 
as the effective $\pi N N\sp{\ast}$ and $\pi \Delta N\sp{\ast}$ interactions,
some of the form factors in Eqs.~(\ref{eq7})-(\ref{eq8})
can be exactly related to the renormalized coupling constants 
\begin{equation}
f\sb{\pi N N\sp{\ast}}(t) =
\frac{\mpi}{\fpi}
\left[
  \frac{1}{2}       F\sp{N N\sp{\ast}}\sb{A,1}(t)
+ \frac{t}{4m\sb{N\sp{\ast}}}F\sp{N N\sp{\ast}}\sb{A,2}(t)
\right],
\label{eq9}
\end{equation}
\begin{eqnarray}
f\sb{\pi \Delta N\sp{\ast}}(t) &=&
\frac{\mpi}{\fpi}
\left[
  F\sp{\Delta N\sp{\ast}}\sb{A,1}(t)
+ (m\sb{N\sp{\ast}} - m\sb{\Delta}) F\sp{\Delta N\sp{\ast}}\sb{A,2}(t)
+ tF\sp{\Delta N\sp{\ast}}\sb{A,3}(t)
\right].
\label{eq10}
\end{eqnarray}

At tree level, all the form factors are reduced to constants.
Because it is difficult to fix all of their value
in the present status of experimental data, we eliminate 
$F\sp{N N\sp{\ast}}\sb{A,2}$ and $F\sp{\Delta N\sp{\ast}}\sb{A,3}$ 
by using the PCAC hypothesis 
$f\sb{\pi N N\sp{\ast}, \pi\Delta N\sp{\ast}}(m\sb{\pi}\sp{2})
\simeq f\sb{\pi N N\sp{\ast}, \pi\Delta N\sp{\ast}}(0)$.
Then, at tree level, we obtain the analogs of Goldberger-Treiman relation 
for the $\pi NN$ interaction,
\begin{equation}
f\sb{\pi N N\sp{\ast}}(\mpi\sp{2})
= \frac{\mpi}{2\fpi}
F\sp{N N\sp{\ast}}\sb{A,1},
\label{eq11}
\end{equation}
and
\begin{equation}
f\sb{\pi \Delta N\sp{\ast}}(\mpi\sp{2})
=
\frac{\mpi}{\fpi} 
\left[
  F\sp{\Delta N\sp{\ast}}\sb{A,1} 
+ (m\sb{N\sp{\ast}}-m\sb{\Delta}) F\sp{\Delta N\sp{\ast}}\sb{A,2}
\right],
\label{eq12}
\end{equation}
respectively.

Furthermore, in this paper we neglect $F\sp{\Delta N\sp{\ast}}\sb{A,2}$
for simplicity.
This would be partly justified by the fact that,
in the case of $\pi N \Delta$ interaction,
the contribution of the $F\sb{A,2}$ term is considerably small compared to
the $F\sb{A,1}$ term~\cite{Kam04}.
The form factor $F\sp{\Delta N\sp{\ast}}\sb{A,4}(t)$
does not appear in our calculation as long as we consider 
the Lagrangian (\ref{eqb7}) for the $\pi \Delta N\sp{\ast}$ interaction.

Using the central values of the $N\sp{\ast}(1440) \rightarrow \pi N$ and
$N\sp{\ast}(1440) \rightarrow \pi \Delta$ decay widths
listed in the particle data table of Ref.~\cite{PDG04},
we obtain 
$f\sb{\pi N N\sp{\ast}}(\mpi\sp{2}) = 0.465$ and
$f\sb{\pi \Delta N\sp{\ast}}(\mpi\sp{2})=1.71$.
As for $f\sb{\pi \Delta N\sp{\ast}}(\mpi\sp{2})$,
we take account of the finite width of $\Delta(1232)$ 
in the same manner as in Ref.~\cite{Gom96}.
Those values lead to $F\sp{N N\sp{\ast}}\sb{A,1} = 0.63$
and $F\sp{\Delta N\sp{\ast}}\sb{A,1} =1.15$, respectively.
\subsubsection{
$N N\sp{\ast} (\pi\pi)\sp{I=0}\sb{S\text{ wave}}$ interaction}
\label{sec2-2-2}
We next consider the $N N\sp{\ast} (\pi\pi)\sp{I=0}\sb{S\text{ wave}}$ 
interaction with the two-pion leg. 
Recently, we have discussed in detail its general
and phenomenological aspects
on the basis of chiral reduction formula~\cite{Kam05-1}. 
We make use of our results also in the present work.

\begin{figure}
\includegraphics[height=3cm]{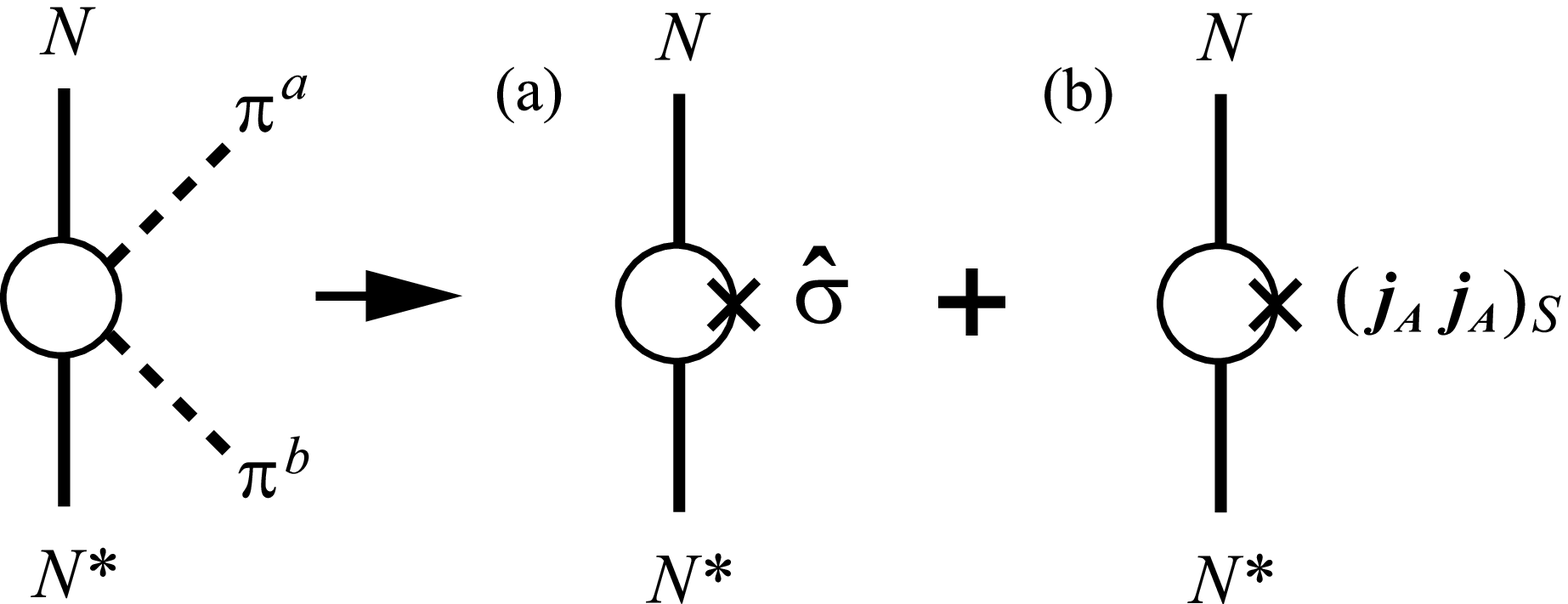}
\caption{
The diagrammatical interpretation for
the chiral reduction of 
the $NN\sp{\ast}(\pi \pi)\sb{S\text{ wave}}\sp{I=0}$ vertex.
The decay amplitude is decomposed into two contributions arising from 
the different origin in the chiral structure: 
one includes (a) the scalar $N\sp{\ast}$-$N$ transition matrix element,
and another includes (b) the $N\sp{\ast}$-$N$ transition matrix element of
the contact, scalar-isoscalar combination of two axial currents.
The former is due to the explicit breaking of chiral symmetry and thus
vanishes in the chiral limit, whereas the latter does not.
}
\label{fig3}
\end{figure}

Using the chiral reduction formula, we find that 
the $NN\sp{\ast}(\pi\pi)\sp{I=0}\sb{S\text{ wave}}$ vertex is
generally described by the following matrix elements of currents and 
density operators (see Fig.~\ref{fig3}).
One is the scalar matrix element of
the $N$-$N\sp{\ast}$ transition [Fig.~\ref{fig3}~(a)], 
which is factorized as
\begin{equation}
\bra{N(p\sp{\prime})}\hat{\sigma}(0)\ket{N\sp{\ast}(p)}
=
-\frac{\sigma\sb{RN}(t)}{\fpi \mpi\sp{2}} 
\bar{u}\sb{N}(p\sp{\prime})u\sb{N\sp{\ast}}(p).
\label{eq13}
\end{equation}
Another is the matrix element of the $N$-$N\sp{\ast}$ transition caused by
the contact, scalar-isoscalar combination of two axial currents 
[Fig.~\ref{fig3}~(b)], which is expressed as 
\begin{eqnarray}
&&
\int d\sp{4}x e\sp{ikx}
\left.
\bra{N(p\sp{\prime})}
T\sp{\ast}\bm{(}\axialjc{a}{\mu}(x)\axialjc{b}{\nu}(0)\bm{)}
\ket{N\sp{\ast}(p)}
\right|\sb{\text{scalar-contact}}
\nonumber\\
&=&
-i \delta\sp{ab}g\sb{\mu\nu}
F\sb{AA}(t) \bar{u}\sb{N}(p\sp{\prime})u\sb{N\sp{\ast}}(p),
\label{eq14}
\end{eqnarray}
where $k$ represents the four-momentum of external pion 
with the isospin index $a$.
Note that $F\sb{AA}(t)$ does not contain any single baryon poles
owing to the definition for the 
$N\sp{\ast}(1440) \rightarrow N(\pi\pi)\sp{I=0}\sb{S\text{ wave}}$ decay
given in Ref.~\cite{PDG04}.
We then obtain the general expression for the
$NN\sp{\ast}(\pi\pi)\sp{I=0}\sb{S\text{ wave}}$ vertex as
\begin{equation}
{\cal M}\sb{NN\sp{\ast}(\pi\pi)\sb{S}} =
\frac{\delta\sp{ab}}{\fpi\sp{2}}
\left[ 
\sigma\sb{RN}(t)
-\frac{t-2\mpi\sp{2}}{2} F\sb{AA}(t)
\right]
\bar{u}\sb{N}(p\sp{\prime})u\sb{N\sp{\ast}}(p).
\label{eq15}
\end{equation}
When $\sigma\sb{RN}(t)$ and $F\sb{AA}(t)$ are constants,
this expression reduces to the result obtained from 
the effective chiral Lagrangian to order $q\sp{2}$~\cite{Ber95}.

\begin{figure}
\includegraphics[height=3cm]{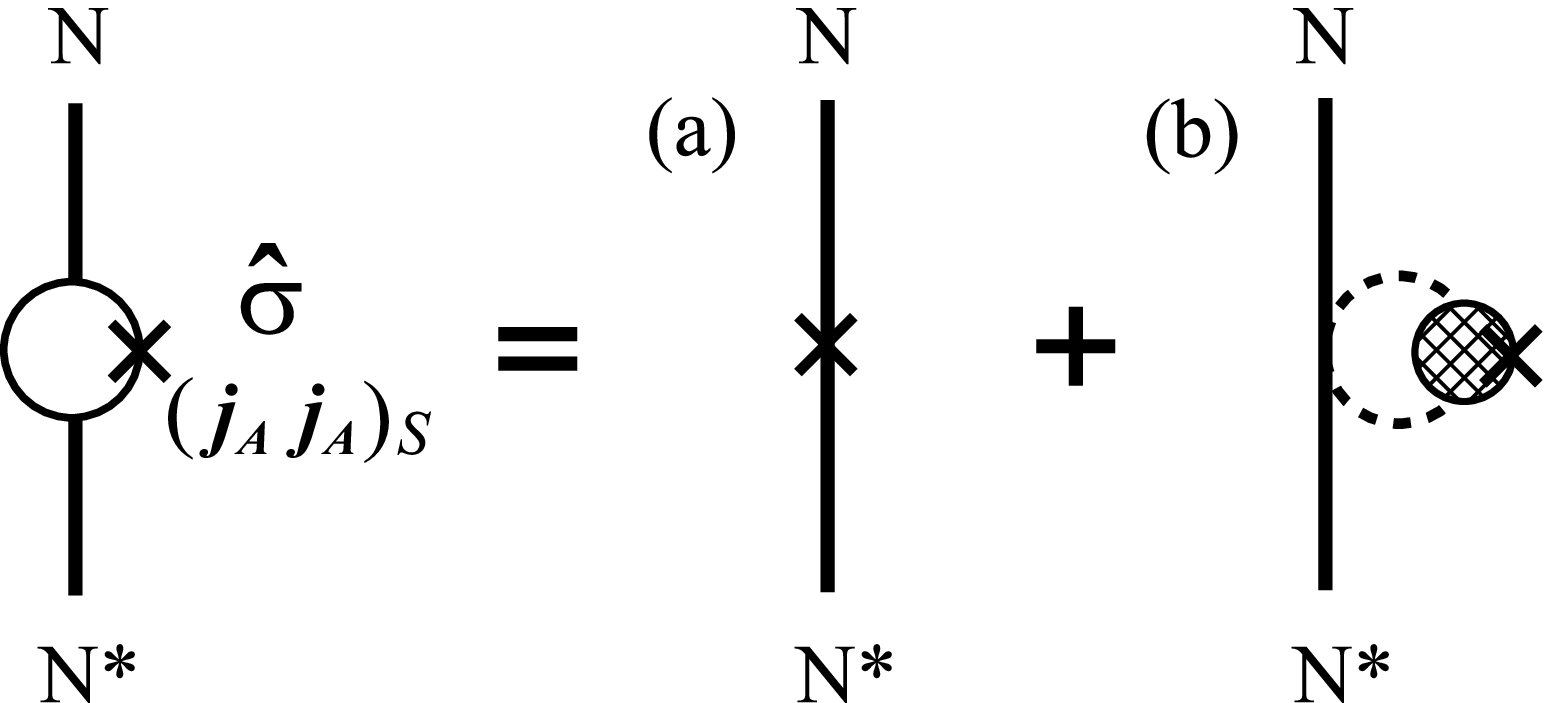}
\caption{
The ``minimal model'' for 
the $NN\sp{\ast}(\pi\pi)\sp{I=0}\sb{S\text{ wave}}$ vertex.
The form factors $\sigma\sb{RN}(t)$ and $F\sb{AA}(t)$ are dominated by 
the processes:
(a) the contact term and (b) the $\pi \pi$ rescattering
together with the contact scalar-isoscalar baryon-pion interaction
(the meshed blob represents the $\pi \pi$ rescattering in $I$=$J$=$0$ channel).
}
\label{fig4}
\end{figure}

Instead of taking $\sigma\sb{RN}(t)$ and $F\sb{AA}(t)$
as constants, we proposed in Ref.~\cite{Kam05-1} 
the ``minimal model'' which explicitly includes
the scalar-isoscalar correlation of two-pions
by considering the $\pi \pi$ rescattering mechanism 
in $I$=$J$=$0$ channel (Fig.~\ref{fig4}).
This model gives the following parameterizations
of $\sigma\sb{RN}(t)$ and $F\sb{AA}(t)$,
\begin{equation}
\sigma\sb{RN}(t) 
\rightarrow
\sigma\sb{RN} \times
\left( 1 + \frac{1}{6}G(t)t\sb{\pi\pi}\sp{I=0}(t) \right),
\label{eq16}
\end{equation}
\begin{equation}
F\sb{AA}(t)
\rightarrow
F\sb{AA} \times
\left( 1+\frac{1}{6}G(t)t\sb{\pi\pi}\sp{I=0}(t) \right),
\label{eq17}
\end{equation}
where $G(t)$ and $t\sb{\pi\pi}\sp{I=0}(t)$ are the pion loop integral and 
$I=0$ $\pi\pi$ rescattering amplitude,
\begin{equation}
G(t) = i\int\frac{d\sp{4}l}{(2\pi)\sp{4}}
\frac{1}{l\sp{2} -\mpi\sp{2} +i\varepsilon}
\frac{1}{(l-P)\sp{2} -\mpi\sp{2} +i\varepsilon}
\label{eq18}
\end{equation}
with $P\sp{2}=t$, and
\begin{equation}
t\sb{\pi\pi}\sp{I=0}(t) = -\frac{6}{\fpi\sp{2}}
\frac{t-\mpi\sp{2}/2}{1+(1/\fpi\sp{2})(t-\mpi\sp{2}/2)G(t)},
\label{eq19}
\end{equation}
respectively~\cite{Kam05-1}.
Based on the dimensional regularization scheme with a renormalization scale
$\mu =1.2$~GeV, the loop integral $G(t)$ can be expressed as
\begin{eqnarray}
G(t) &=& 
\frac{1}{(4\pi)\sp{2}}
\left(-1+\ln\frac{\mpi\sp{2}}{\mu\sp{2}}
+\sigma\ln\frac{1+\sigma}{1-\sigma}-i\pi\sigma
\right)
\label{eq20}
\end{eqnarray}
for $t > 4\mpi\sp{2}$,
\begin{eqnarray}
G(t) &=&
\frac{1}{(4\pi)\sp{2}}
\left(
-1 + \ln\frac{\mpi\sp{2}}{\mu\sp{2}}
+\sigma \ln \frac{\sigma +1}{\sigma -1}
\right)
\label{eq21}
\end{eqnarray}
for $t < 0$, and
\begin{eqnarray}
G(t) &=&
\frac{1}{(4\pi)\sp{2}}
\left(
-1 + \ln\frac{\mpi\sp{2}}{\mu\sp{2}} + \sigma (\pi -2\arctan\sigma) 
\right)
\label{eq22}
\end{eqnarray}
for $ 0 < t <  4\mpi\sp{2}$,
where $\sigma=\sqrt{|1-(4\mpi\sp{2}/t)|}$.

Using Eqs.~(\ref{eq15})-(\ref{eq17}) and performing the phase space integral, 
we obtain the following expression for 
the $N\sp{\ast}(1440)\rightarrow N(\pi\pi)\sp{I=0}\sb{S\text{ wave}}$ decay
width
\begin{equation}
\Gamma\sb{NN\sp{\ast}(\pi\pi)\sb{S}}
= 
  \alpha (c\sb{1}\sp{\ast})\sp{2}
+ \beta  (c\sb{2}\sp{\ast})\sp{2}
+ \gamma c\sb{1}\sp{\ast}c\sb{2}\sp{\ast},
\label{eq23}
\end{equation}
where $c\sb{1}\sp{\ast}=-\sigma\sb{RN}/(2\mpi\sp{2})$
and $c\sb{2}\sp{\ast}=F\sb{AA}/2$.
The numerical value of the coefficients 
$\alpha$, $\beta$, and $\gamma$ is
$\alpha=1.199\times 10\sp{-3}~\text{GeV}\sp{3}$,
$\beta=14.06\times 10\sp{-3}~\text{GeV}\sp{3}$,
and $\gamma=7.754\times 10\sp{-3}~\text{GeV}\sp{3}$, respectively.
We note that these values are different
from those of Refs.~\cite{Ber95,Alv98}.
This difference arises from the $\pi\pi$ rescattering
and the relativistic effects.
Both $c\sp{\ast}\sb{1}$ and $c\sp{\ast}\sb{2}$
are not fixed by the 
$N\sp{\ast}(1440) \rightarrow N(\pi\pi)\sb{S\text{ wave}}\sp{I=0}$ decay width,
and we can take any value on the 
ellipse~(\ref{eq23}) on the $c\sb{1}\sp{\ast}$-$c\sb{2}\sp{\ast}$ plane.
\subsection{Cut-off factor}
\label{sec2-3}
We shall introduce the cut-off factors, 
which stem from the finite size of hadrons,
for each vertex in a phenomenological manner.
In Ref.~\cite{Kam04} we treated all hadrons as the point-like particles.
They are, however, the bound states of quarks and gluons
and thus have the finite size.
Although in principle those effects should be directly derived from QCD,
it is very difficult and still an open question
which deserves the theoretical challenges.
Instead, we consider those effects phenomenologically
according to the discussions of Ref.~\cite{Pea91}.
We attach the four-dimensional cut-off factors,
\begin{equation}
f\sb{\alpha}(p\sb{\alpha}\sp{2}) = 
\frac{\Lambda\sb{\alpha}\sp{4}}
{\Lambda\sb{\alpha}\sp{4} + (p\sb{\alpha}\sp{2} - m\sb{\alpha}\sp{2})\sp{2}},
\label{eq24}
\end{equation}
to each leg $\alpha$ of the vertex 
(where we assume $n\sb{\alpha}=1$~\cite{Pea91}).
Here $p\sb{\alpha}$ and $m\sb{\alpha}$ are the four-momentum and mass
of the particle corresponding to the leg $\alpha$, respectively. 
The cut-off factor $f\sb{\alpha}$ is normalized to one on the mass-shell,
and thus exhibits its effect only for the internal lines.
There are five cut-off parameters:
$\Lambda\sb{N}$, $\Lambda\sb{\Delta}$, and $\Lambda\sb{N\sp{\ast}}$
for the intermediate baryons,
and $\Lambda\sb{\pi}$ and $\Lambda\sb{\rho}$
for the pion and $\rho$ meson poles
appearing in Figs.~\ref{fig2}(a) and~\ref{fig2}(d).
In this paper we assume
$\Lambda\sb{N} = \Lambda\sb{\Delta} 
= \Lambda\sb{N\sp{\ast}} \equiv \Lambda\sb{B}$
and 
$\Lambda\sb{\pi} = \Lambda\sb{\rho} \equiv \Lambda\sb{M}$.
\section{Results and Discussions}
\label{sec3}
\begin{figure}
\includegraphics[width=5cm]{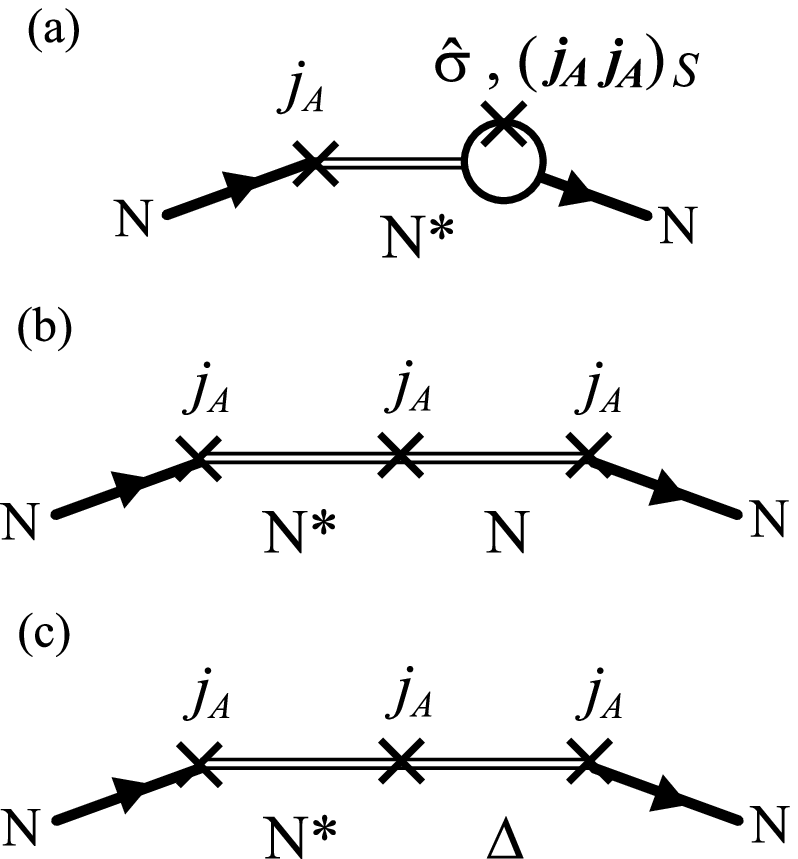}
\caption{
The diagrams including $N\sp{\ast}(1440)$ 
taken into account in the present paper.
All crossed versions of these diagrams are also considered.
}
\label{fig5}
\end{figure}
Putting together all diagrams introduced
in our previous study (Fig.~\ref{fig2}) and 
this work (Fig.~\ref{fig5}),
we calculate the $\pi N \rightarrow \pi \pi N$ total cross sections 
up to $T\sb{\pi}=620$~MeV. 
We also calculate the $\pi\sp{0}\pi\sp{0}$ and $\pi\sp{0}n$ 
invariant mass distributions
for the $\pi\sp{-} p \rightarrow \pi\sp{0} \pi\sp{0} n$ reactions
whose experimental data were reported by CBC~\cite{Pra04}.
We especially try to discuss
the contributions of $N\sp{\ast}(1440)$ around its mass-shell energy.
\subsection{Fixing parameters}
\label{sec3-1}
Before showing our results,
we mention how to fix the parameters:
the cut-off parameters $\Lambda\sb{B}$ and $\Lambda\sb{M}$,
and the coupling constants of the $\pi \Delta \Delta$ 
and $\rho N \Delta$ interactions 
(denoted as $f\sb{\pi \Delta \Delta}$ and $f\sb{\rho N \Delta}$, respectively),
and the parameters $c\sp{\ast}\sb{1}$ and $c\sp{\ast}\sb{2}$ 
which characterize $N N\sp{\ast} (\pi \pi)\sp{I=0}\sb{S\text{ wave}}$ 
interaction.

Because the value of $f\sb{\pi \Delta \Delta}$ and $f\sb{\rho N \Delta}$
can not be directly determined by the two-body decay width,
they are usually extracted from the data using meson 
exchange model, or estimated by using theoretical approach 
such as the quark model and QCD sum rule.
Many reports on their value
fall into the range of $0.4 \alt f\sb{\pi \Delta \Delta} \alt 0.8$
and $3.5 \alt f\sb{\rho N \Delta} \alt 7.8$~(see \cite{Kam04} and references
therein).
For instance, the quark model relations lead to 
$f\sb{\pi \Delta \Delta}=0.8$ and
$f\sb{\rho N \Delta}=5.5$~\cite{Bro75,Oh05}. 

In Ref.~\cite{Kam04}, 
it is found that, at least in the low energy region up to $T\sb{\pi}=400$ MeV,
the influences of $\pi \Delta \Delta$ and $\rho N \Delta$ interactions 
are negligible for the $\pi\sp{-} p \rightarrow \pi\sp{+} \pi\sp{-} n$ 
and $\pi\sp{-} p \rightarrow \pi\sp{0} \pi\sp{0} n$ channels.
The $N N\sp{\ast}(\pi \pi)\sp{I=0}\sb{S\text{ wave}}$ interaction
(i.e. $c\sp{\ast}\sb{1}$ and $c\sp{\ast}\sb{2}$), on the other hand,
gives a negligible contribution to other three channels, i.e.
$\pi\sp{\pm} p \rightarrow \pi\sp{\pm} \pi\sp{0} p$
and $\pi\sp{+} p \rightarrow \pi\sp{+} \pi\sp{+} n$~\cite{Jen97}.
This fact allows us to consider 
$(f\sb{\rho N \Delta},f\sb{\pi \Delta \Delta})$
and $(c\sp{\ast}\sb{1},c\sp{\ast}\sb{2})$
separately in the parameter fixing. 
Thus we first try to investigate the influences of
$\Lambda\sb{B}$, $\Lambda\sb{M}$, 
$f\sb{\pi \Delta \Delta}$, and $f\sb{\rho N \Delta}$
on the total cross sections of
$\pi\sp{\pm} p \rightarrow \pi\sp{\pm} \pi\sp{0} p$
and $\pi\sp{+} p \rightarrow \pi\sp{+} \pi\sp{+} n$ channels,
and to estimate the value of these parameters.
\subsubsection{Cut-off dependence}
\label{sec3-1-1}
\begin{figure}
\includegraphics[width=5cm]{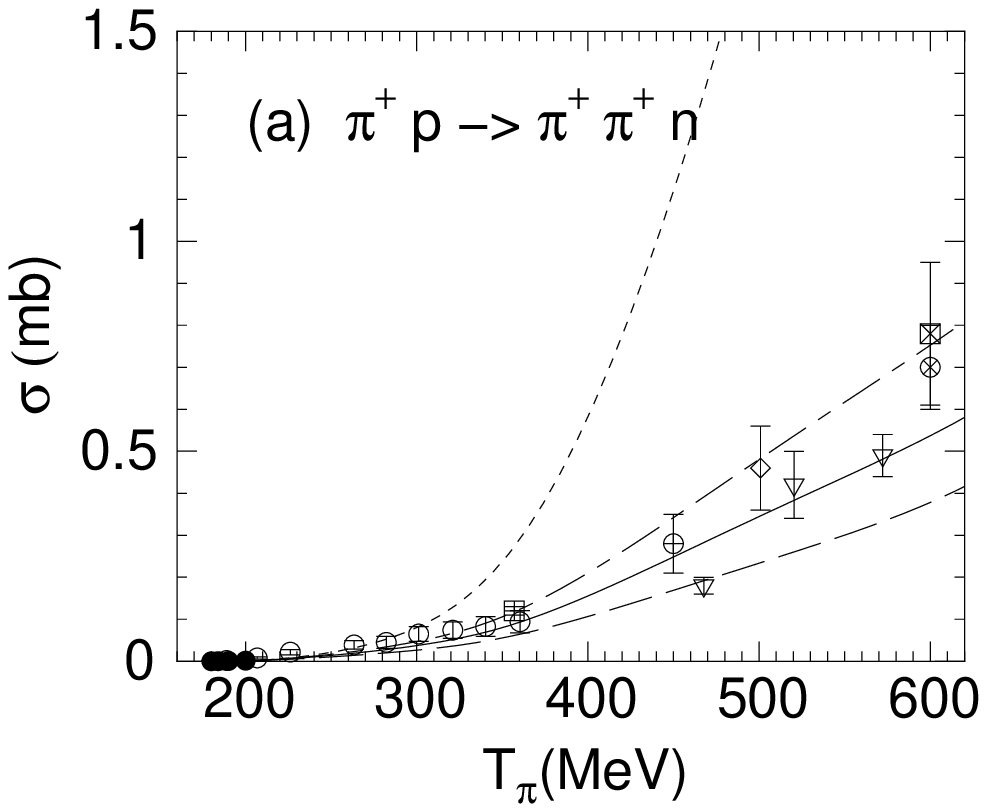}
\includegraphics[width=5cm]{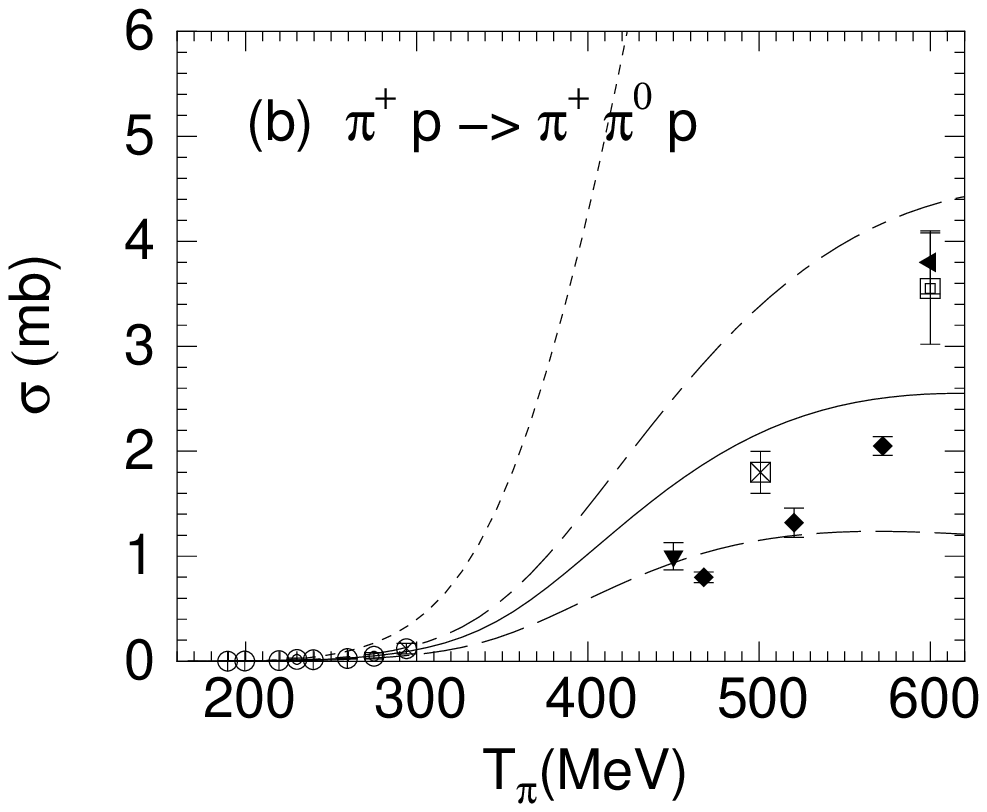}
\includegraphics[width=5cm]{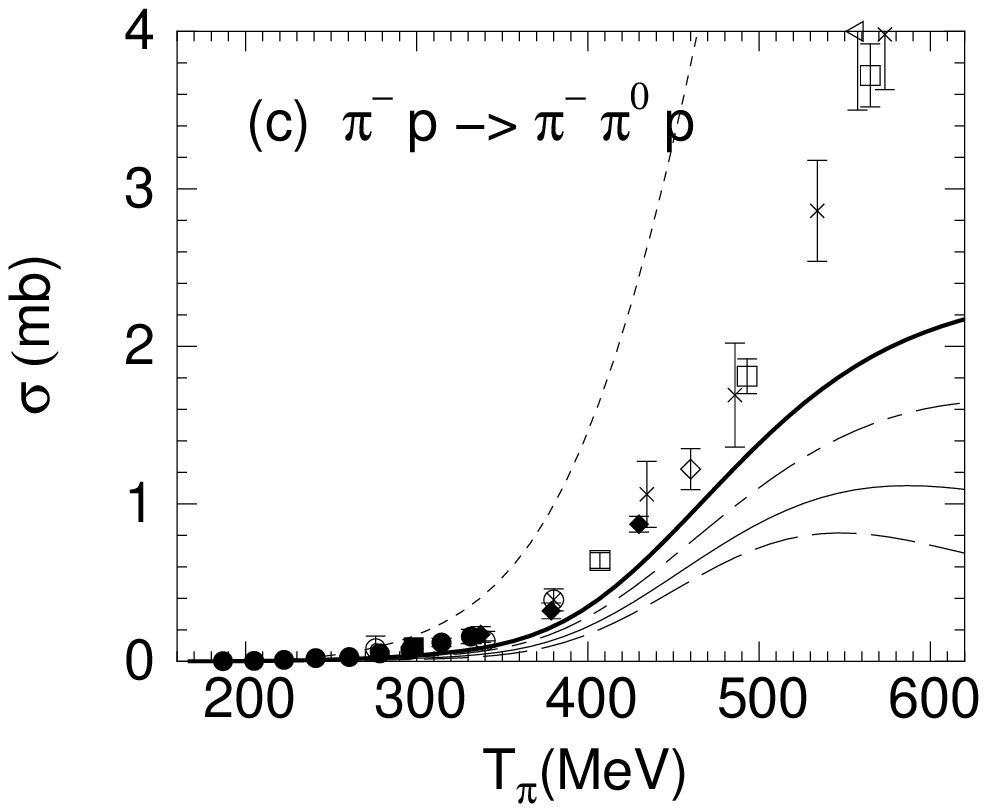}
\caption{
The cut-off dependence of the total cross sections.
The results are shown for 
the cases of $\Lambda\sb{B}=\Lambda\sb{M}=750$~MeV (dashed line),
$\Lambda\sb{B}=\Lambda\sb{M}=850$~MeV (solid line), and
$\Lambda\sb{B}=\Lambda\sb{M}=950$~MeV (dashed-dotted line).
The results without considering the cut-off factors are shown by 
the dotted line.
The thick solid line in (c) corresponds to the case of
$\Lambda\sb{B}=850$~MeV and $\Lambda\sb{M}=2\Lambda\sb{B}=1700$~MeV.
The data are from 
Refs.~\cite{Sev91,Ker90,Kir62,Bar62,Deb65,Bow70,Poi66-1,New63,Bat75,
Poc94,Arm72,Sax70,Jon74,Blo62,Dol74,Deb69,Poi66-2,Bur65,Vit64,Bro71-2,
Ker89-2,Bar61}.
}
\label{fig6}
\end{figure}
In Fig.~\ref{fig6} we show the cut-off dependence of the total cross sections.
In this calculation, we include all diagrams but 
Fig.~\ref{fig5}(a) which includes
the $N N\sp{\ast} (\pi \pi)\sp{I=0}\sb{S\text{ wave}}$ vertex,
and take $f\sb{\pi \Delta \Delta}=0.4$ and
$f\sb{\rho N \Delta}=7.8$ which are in the range explained 
above\footnote{The reason for taking these value will be discussed later.}.

First of all, if we do not take account of 
the cut-off factor~(\ref{eq24}) for each leg of the vertices, 
i.e. if we treat the hadrons as point-like particle, 
the resulting total cross sections obviously overshoot 
the experimental data above $T\sb{\pi}\sim 300$~MeV.

We choose three typical values
$\Lambda\sb{B}=\Lambda\sb{M}=750,\ 850,\ 950$~MeV,
and calculate the cross sections for each value.
As shown in Fig.~\ref{fig6}, we can see clear
difference among these results above $T\sb{\pi}\sim 300$~MeV.
For the $\pi\sp{+} p \rightarrow \pi\sp{+} \pi\sp{+} n$ channel,
the case of $\Lambda\sb{B}=\Lambda\sb{M}=850$~MeV seems most appropriate.
The difference is obviously seen in
the results for the $\pi\sp{+} p \rightarrow \pi\sp{+} \pi\sp{0} p$ channel.
The case of $\Lambda\sb{B}=\Lambda\sb{M}=950$~MeV
is quite different from the data compared to the other two cases.
For the $\pi\sp{-} p \rightarrow \pi\sp{-} \pi\sp{0} p$ channel,
all results somewhat underestimate the data.

Next we consider several cases of $\Lambda\sb{B}{\not =}\Lambda\sb{M}$:
$\Lambda\sb{M}=0.5\Lambda\sb{B},\ 1.5\Lambda\sb{B},\ 2\Lambda\sb{B}$
for each value of $\Lambda\sb{B}$.
As for the $\pi\sp{+} p \rightarrow \pi\sp{+} \pi\sp{+} n$
and $\pi\sp{+} p \rightarrow \pi\sp{+} \pi\sp{0} p$ channels
the results change only within 10 \% 
when we vary the value of $\Lambda\sb{M}$,
whereas the $\pi\sp{-} p \rightarrow \pi\sp{-} \pi\sp{0} p$ channel
is sensitive to the variation of $\Lambda\sb{M}$. 
We find that the larger value of $\Lambda\sb{M}$ for each $\Lambda\sb{B}$
seems appropriate:
the thick solid line in Fig.~\ref{fig6}(c) for
$\Lambda\sb{B}=850$~MeV and $\Lambda\sb{M}=2\Lambda\sb{B}=1700$~MeV
(this line should be compared to the thin solid line with 
$\Lambda\sb{B}=\Lambda\sb{M}=850$~MeV).

In the following discussions, we take 
$\Lambda\sb{B}=850$~MeV and $\Lambda\sb{M}=1700$~MeV.
\subsubsection{The $\pi \Delta \Delta$ and $\rho N \Delta$ interactions}
\label{sec3-1-2} 
\begin{figure}
\includegraphics[width=5cm]{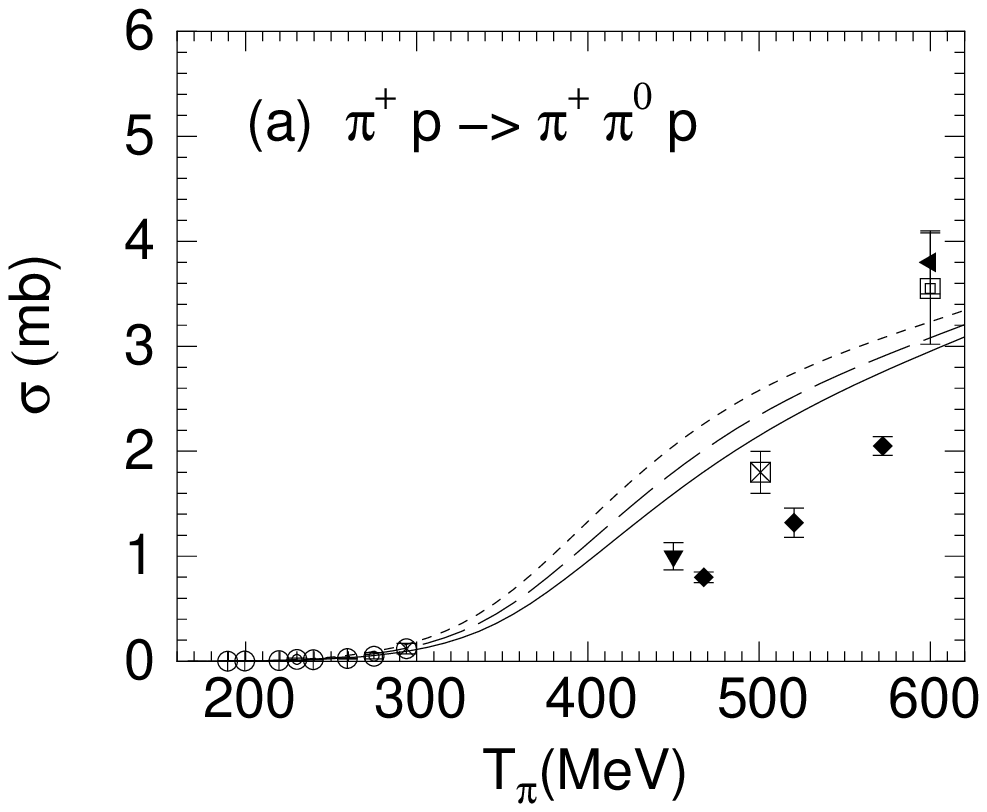}
\hspace{1cm}
\includegraphics[width=5cm]{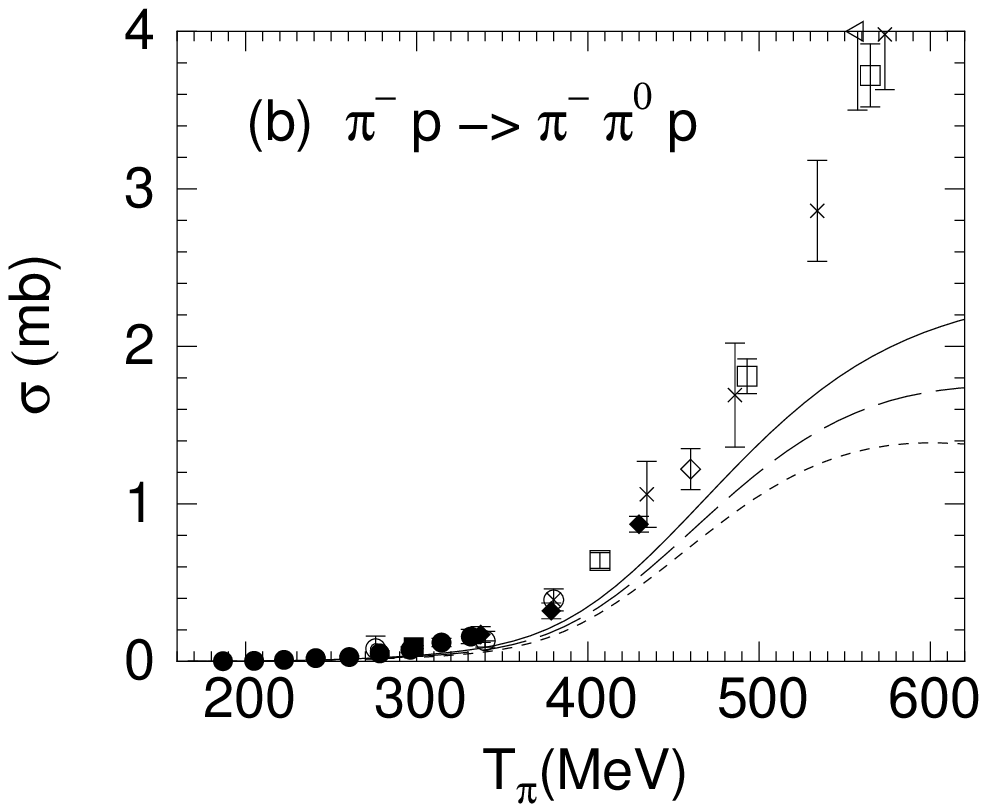}
\caption{
The dependence of numerical results
on $f\sb{\pi \Delta \Delta}$ and $f\sb{\rho N \Delta}$: 
(a) the $\pi\sp{+} p \rightarrow \pi\sp{+} \pi\sp{0} p$ channel
with 
$(f\sb{\pi \Delta \Delta},f\sb{\rho N \Delta})=(0.4,7.8)$ (solid line),
$(f\sb{\pi \Delta \Delta},f\sb{\rho N \Delta})=(0.6,7.8)$ (dashed line),
and
$(f\sb{\pi \Delta \Delta},f\sb{\rho N \Delta})=(0.8,7.8)$ (dotted line),
and
(b) the $\pi\sp{-} p \rightarrow \pi\sp{-} \pi\sp{0} p$ channel
with 
$(f\sb{\pi \Delta \Delta},f\sb{\rho N \Delta})=(0.4,7.8)$ (solid line),
$(f\sb{\pi \Delta \Delta},f\sb{\rho N \Delta})=(0.4,5.7)$ (dashed line),
and
$(f\sb{\pi \Delta \Delta},f\sb{\rho N \Delta})=(0.4,3.5)$ (dotted line).
The contribution of Fig.~\ref{fig5}(a) is not taken into account.
We use $\Lambda\sb{B}=850$~MeV and $\Lambda\sb{M}=1700$~MeV.
The data are the same as in Fig.~\ref{fig6}.
}
\label{fig7}
\end{figure}
Here we try to examine how the total cross section changes
according to the variation of 
$f\sb{\pi \Delta \Delta}$ and $f\sb{\rho N \Delta}$
in their allowed range.
For the $\pi\sp{+} p \rightarrow \pi\sp{+} \pi\sp{+} n$ channel,
the change is small, and the results still compatible with the data.
For the $\pi\sp{+} p \rightarrow \pi\sp{+} \pi\sp{0} p$ channel,
the influence of the $\rho N \Delta$ interaction can be hardly seen
for $3.5 \alt f\sb{\rho N \Delta} \alt 7.8$, 
whereas the $\pi \Delta \Delta$ interaction
generates visible change on this channel [Fig.~\ref{fig7}(a)].
The lower value of $f\sb{\pi \Delta \Delta}$
seems appropriate to the data.
In contrast, for the variation of $f\sb{\rho N \Delta}$
the change of the 
$\pi\sp{-} p \rightarrow \pi\sp{-} \pi\sp{0} p$ channel
is clearly seen [Fig.~\ref{fig7}(b)]
but is negligible for that of $f\sb{\pi \Delta \Delta}$.
The higher value of $f\sb{\rho N \Delta}$ seems appropriate to the data.

This is why we used $f\sb{\pi \Delta \Delta}=0.4$
and $f\sb{\rho N \Delta} =7.8$ in the discussions of cut-off dependence
of the total cross sections.
Also in the following we continue to use these values for 
$f\sb{\pi \Delta \Delta}$ and $f\sb{\rho N \Delta}$.
\subsubsection{The dependence on $c\sp{\ast}\sb{1}$ and $c\sp{\ast}\sb{2}$}
\label{sec3-1-3}
\begin{figure}
\includegraphics[width=5cm]{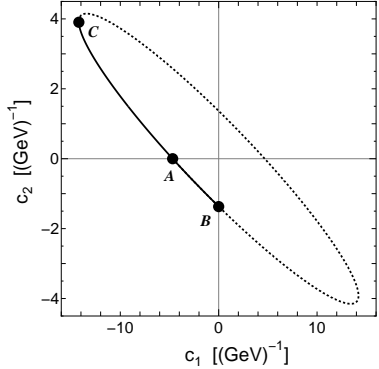}
\caption{
The ellipse (23) representing the allowed values of $c\sp{\ast}\sb{1}$ and 
$c\sp{\ast}\sb{2}$. 
Here $\Gamma\sb{NN\sp{\ast}(\pi\pi)\sb{S}}=26.25$~MeV
is used, which is the central value
listed in the particle data table~\cite{PDG04}.
The point $A$, which is 
$(c\sp{\ast}\sb{1},c\sp{\ast}\sb{2})=(-4.68,0)~\rm{GeV}\sp{-1}$ 
in our case, corresponds to the point often used 
in the study of hadron reactions.
The points $B$ and $C$ correspond to
$(c\sp{\ast}\sb{1},c\sp{\ast}\sb{2})=(0,-1.37)$~GeV$\sp{-1}$ and
$(c\sp{\ast}\sb{1},c\sp{\ast}\sb{2})=(-14.2,3.92)$~GeV$\sp{-1}$, respectively.
See the text for the meanings of the solid and dotted parts of this ellipse.
}
\label{fig8}
\end{figure}
As mentioned in Subsec.~\ref{sec2-2}, 
the values of $c\sp{\ast}\sb{1}$ and $c\sp{\ast}\sb{2}$ can not be fixed
by the relation~(\ref{eq23}) with the decay width of
$N\sp{\ast}(1440) \rightarrow N (\pi \pi)\sp{I=0}\sb{S\text{ wave}}$,
which just represents the ellipse on 
the $c\sp{\ast}\sb{1}$-$c\sp{\ast}\sb{2}$ plane as shown in Fig.~\ref{fig8}.
Although the case of $c\sp{\ast}\sb{2}=0$ and $c\sp{\ast}\sb{1} < 0$
(the point $A$ in Fig.~\ref{fig8})
is conventionally used in the phenomenological models
of hadron reactions~(e.g. see Refs.~\cite{Ose85,Gom96}),
it was first pointed out in Ref.~\cite{Ber95} that 
the $c\sp{\ast}\sb{2}$ contribution should be also considered 
in view of the general chiral effective Lagrangian.
Several attempts have been recently performed to determine the values of 
$c\sp{\ast}\sb{1}$ and $c\sp{\ast}\sb{2}$~\cite{Alv98,Her99}.

Calculating the total cross sections 
for all five channels up to $T\sb{\pi}=620$ MeV 
with allowed values of $c\sp{\ast}\sb{1}$ and $c\sp{\ast}\sb{2}$,
we find that the appropriate values
would be on the solid curve in Fig.~\ref{fig8}.
The values on the dotted-curve generate
obvious disagreement in the total cross section,
which is consistent with other studies
of the $\pi N \rightarrow \pi \pi N$ reaction~\cite{Jen97}.

\begin{figure}
\includegraphics[width=5cm]{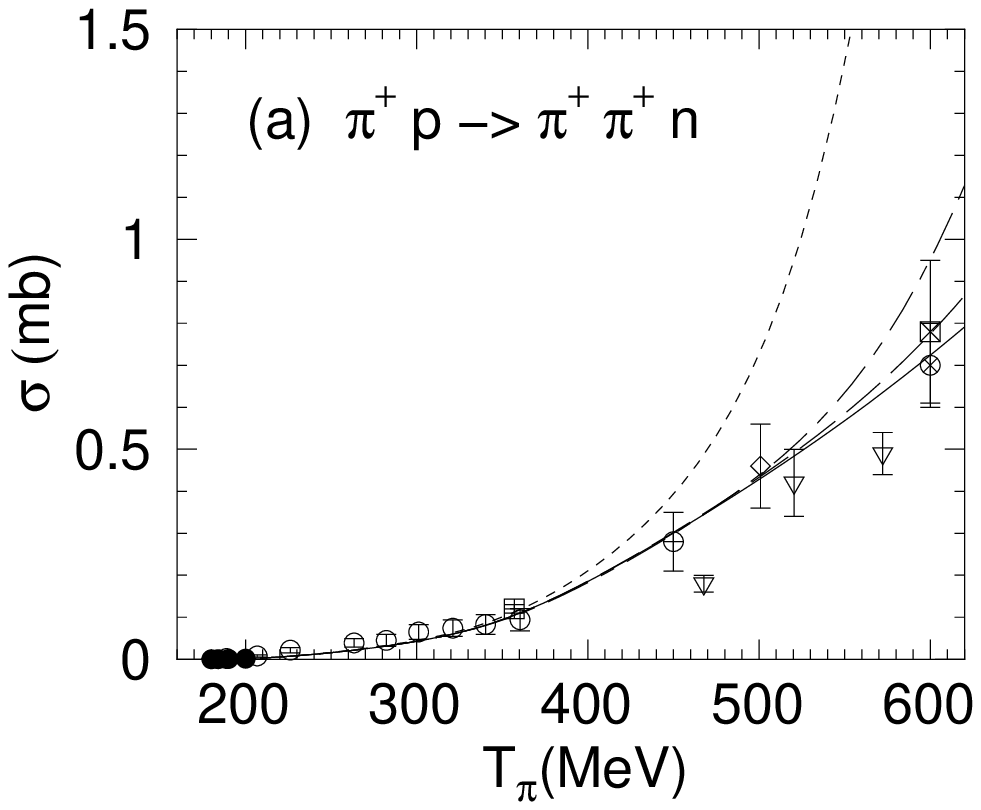}
\includegraphics[width=5cm]{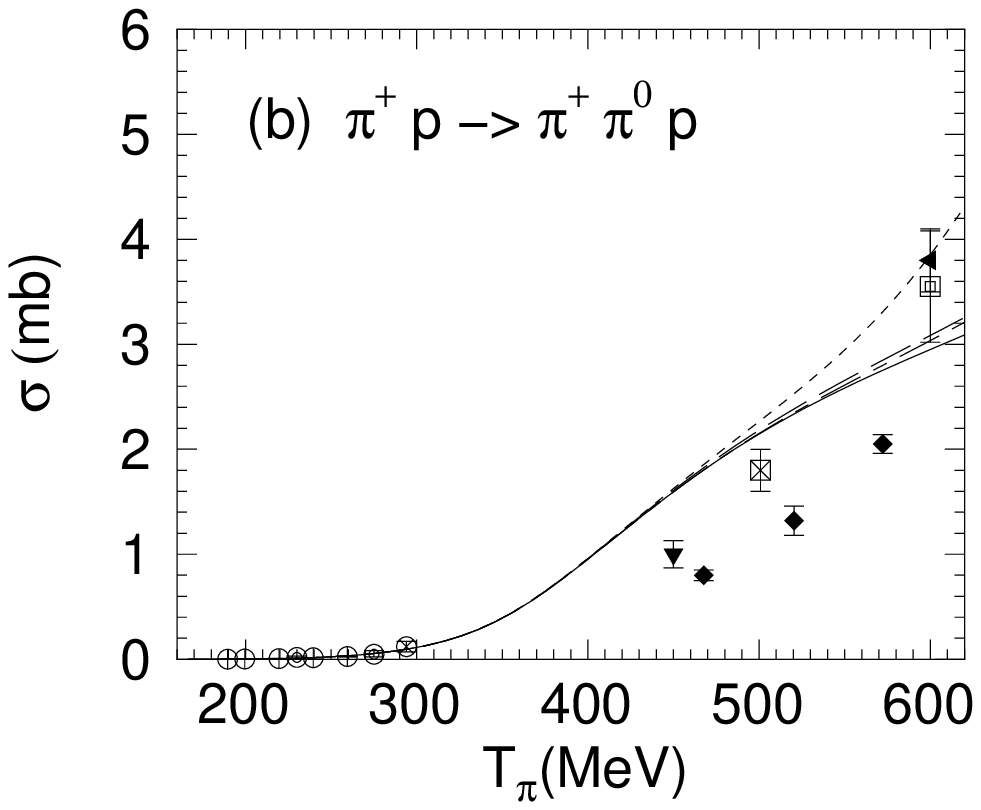}
\includegraphics[width=5cm]{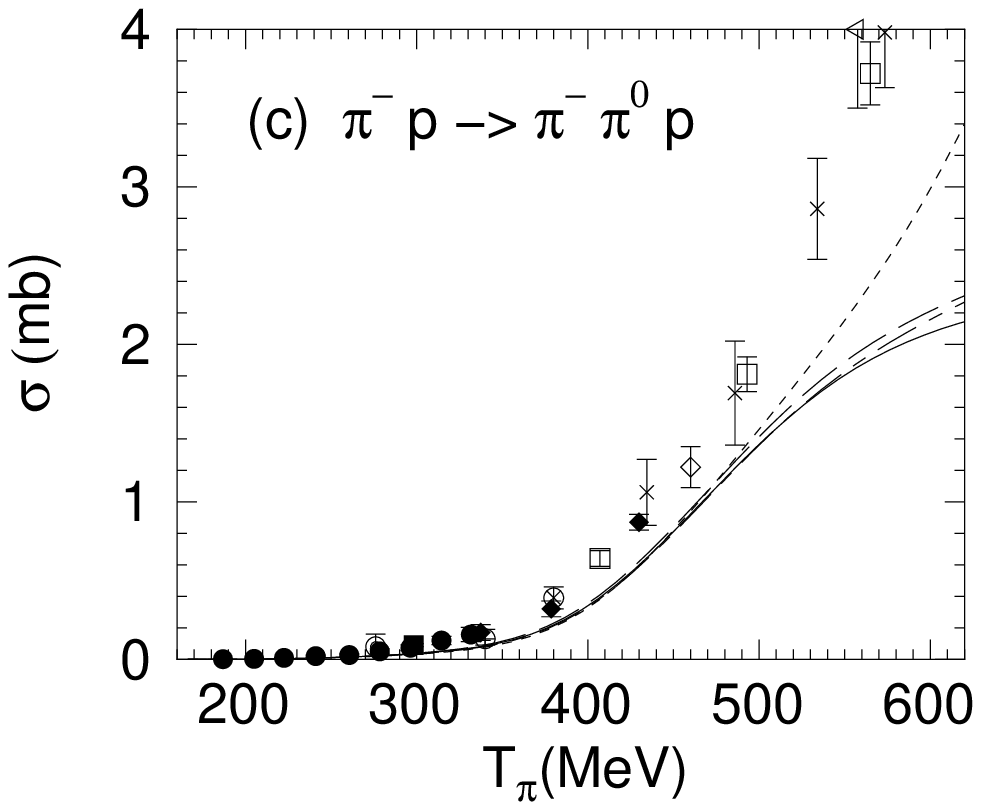}
\includegraphics[width=5cm]{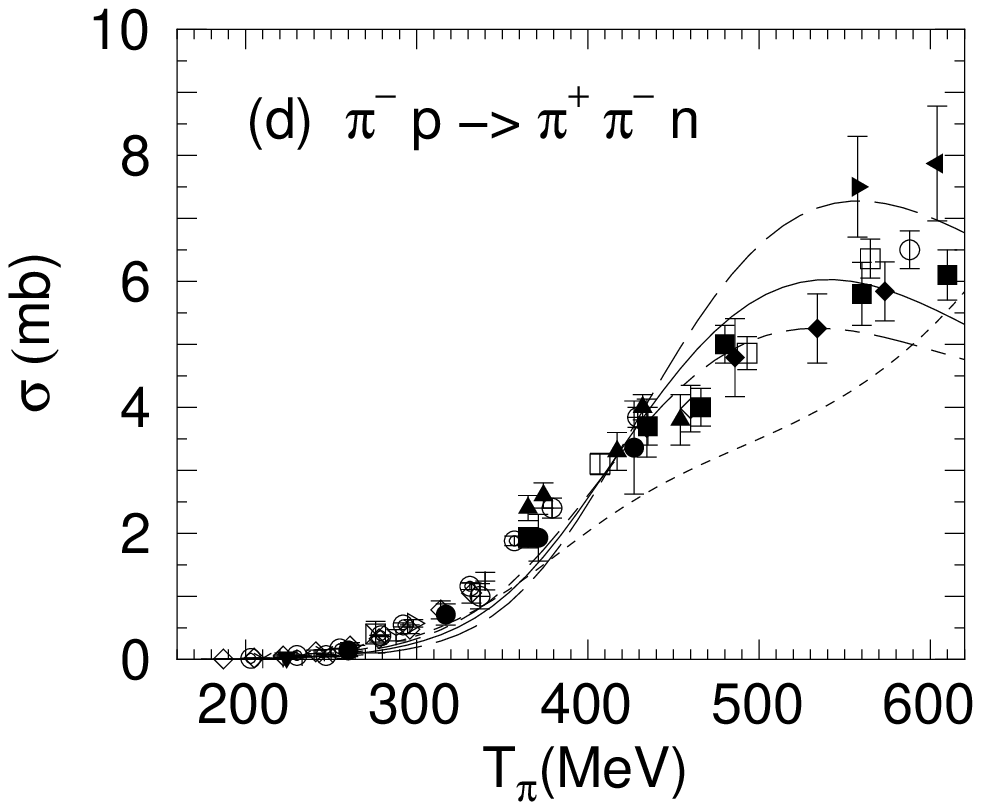}
\includegraphics[width=5cm]{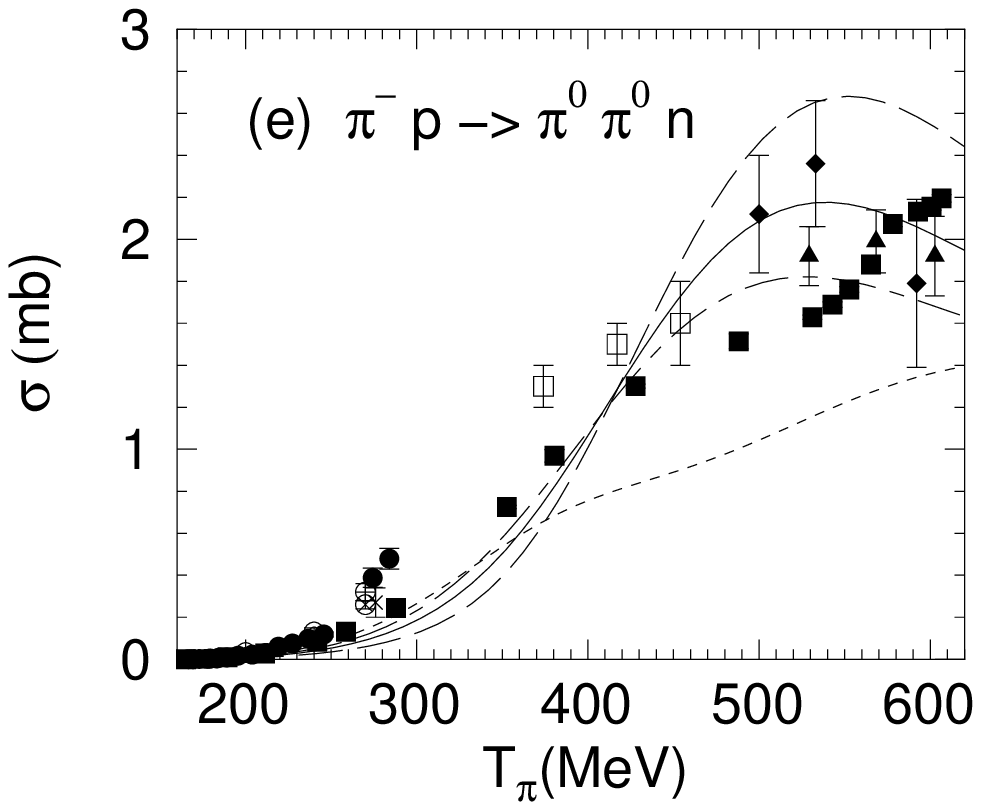}
\caption{
The $c\sp{\ast}\sb{1}$ and $c\sp{\ast}\sb{2}$ dependence of
the total cross sections. 
The solid line is the result with $(c\sp{\ast}\sb{1},c\sp{\ast}\sb{2})$
at the point $A$ on the ellipse,
whereas the dashed and dotted lines are at the point $B$ and $C$, 
respectively.
The dotted-dashed line is for 
$(c\sp{\ast}\sb{1},c\sp{\ast}\sb{2})=(-8.0,1.08)$~GeV$\sp{-1}$.
The data are the same as in Fig.~\ref{fig8} for (a)-(c),
whereas are taken from 
Refs.~\cite{Per60,Dea61,Bjo80,Sax70,Bla70,Jon74,Ker89-1,Bat65,
Blo63,Blo62,Dol74,Deb69,Poi66-2,Bur65,Vit64,Bar64,Bro71-2,Kir63,
Bel80,Kra75,Bun77,Bel78,Low91,Bul69,Cen67}
for (d) and (e).
The CBC data are plotted as the filled square in (e).
}
\label{fig9}
\end{figure}
In Fig.~\ref{fig9}, we show the total cross sections calculated with 
several values of $c\sp{\ast}\sb{1}$ and $c\sp{\ast}\sb{2}$ 
on the solid curve in Fig.~\ref{fig8}.
The recent CBC data for the $\pi\sp{-} p \rightarrow \pi\sp{0} \pi\sp{0} n$
channel are plotted as the filled square in Fig.~\ref{fig9}~(e).
We notice the CBC data exhibit a bump below the $N\sp{\ast}(1440)$ energy, 
i.e. 350 $\alt T\sb{\pi} \alt$ 500 MeV,
and this bump is called the ``shoulder'' in Ref.~\cite{Pra04}.

For the $\pi\sp{\pm} p \rightarrow \pi\sp{\pm} \pi\sp{0} p$ channels,
the results show only a small increase in the total cross section 
above $T\sb{\pi}=500$~MeV except for $c\sp{\ast}\sb{1}$ and $c\sp{\ast}\sb{2}$
around the point $C$ [see Figs.~\ref{fig9}(b) and~\ref{fig9}(c)].
Below $T\sb{\pi}=500$~MeV, however,
the $NN\sp{\ast}(\pi\pi)\sp{I=0}\sb{S\text{ wave}}$ interaction
has no influence on the total cross section.
This result is consistent with other studies~\cite{Jen97,Kam04}.

However,
the $\pi\sp{-} p \rightarrow \pi\sp{+} \pi\sp{-} n$ and
$\pi\sp{-} p \rightarrow \pi\sp{0} \pi\sp{0} n$ channels
are sensitive to the variation of 
$c\sp{\ast}\sb{1}$ and $c\sp{\ast}\sb{2}$
[Figs.~\ref{fig9}(d) and~\ref{fig9}(e)].
This fact is expected from other studies at 
low energy~\cite{Ber95,Jen97}.
The results with $c\sp{\ast}\sb{1}$ and $c\sp{\ast}\sb{2}$
near the points $B$ and $C$ are obviously incompatible with
the experimental data.
The preferable value of $c\sp{\ast}\sb{1}$ 
seems to be found between the points $A$ and $C$
in view of the data.

The $\pi\sp{+} p \rightarrow \pi\sp{+} \pi\sp{+} n$ channel 
is also sensitive to 
the variation of $c\sp{\ast}\sb{1}$ and $c\sp{\ast}\sb{2}$
[see Fig.~\ref{fig9}(a)].
The total cross section
above $T\sb{\pi} = 400$~MeV is increased by
the contribution of
the process Fig.~\ref{fig5}(a):
the lower the value of $c\sb{1}\sp{\ast}$ becomes, 
the larger its increase is.
Because this channel is, however, already reproduced well without
$N\sp{\ast}(1440)$, accurate data of 
the $\pi\sp{+} p \rightarrow \pi\sp{+} \pi\sp{+} n$ reaction above 
$T\sb{\pi}=400$~MeV will strongly constrain the value of 
$c\sp{\ast}\sb{1}$ and $c\sp{\ast}\sb{2}$ 
not to disturb this present status.
From the present situation of 
$\pi\sp{+} p \rightarrow \pi\sp{+} \pi\sp{+} n$ data, 
we find that $c\sp{\ast}\sb{1} \geq -8.0~\text{GeV}\sp{-1}$ is preferable
in our model.

It is worth noting that, in contrast to other three channels, 
our results for 
the $\pi\sp{-} p \rightarrow \pi\sp{+} \pi\sp{-} n$ and
 $\pi\sp{-} p \rightarrow \pi\sp{0} \pi\sp{0} n$ channels
show peak structure around the $N\sp{\ast}(1440)$ energy 
corresponding to $T\sb{\pi} \sim 480$~MeV, 
except for those with extremely low values of 
$c\sp{\ast}\sb{1}$ near the point $C$.
This feature should be compared with the CBC data 
for the $\pi\sp{-} p \rightarrow \pi\sp{0} \pi\sp{0} n$ total cross section
which shows the shoulder at the $N\sp{\ast}(1440)$ energy.

Considering all above results, we choose
$c\sp{\ast}\sb{1} = -8.0~\text{GeV}\sp{-1}$ 
(this leads to $c\sp{\ast}\sb{2}=1.08~\text{GeV}\sp{-1}$)
as a plausible value for this parameter.
\subsection{Each contribution of the diagram including $N\sp{\ast}(1440)$}
\label{sec3-2}
\begin{figure}
\includegraphics[width=5cm]{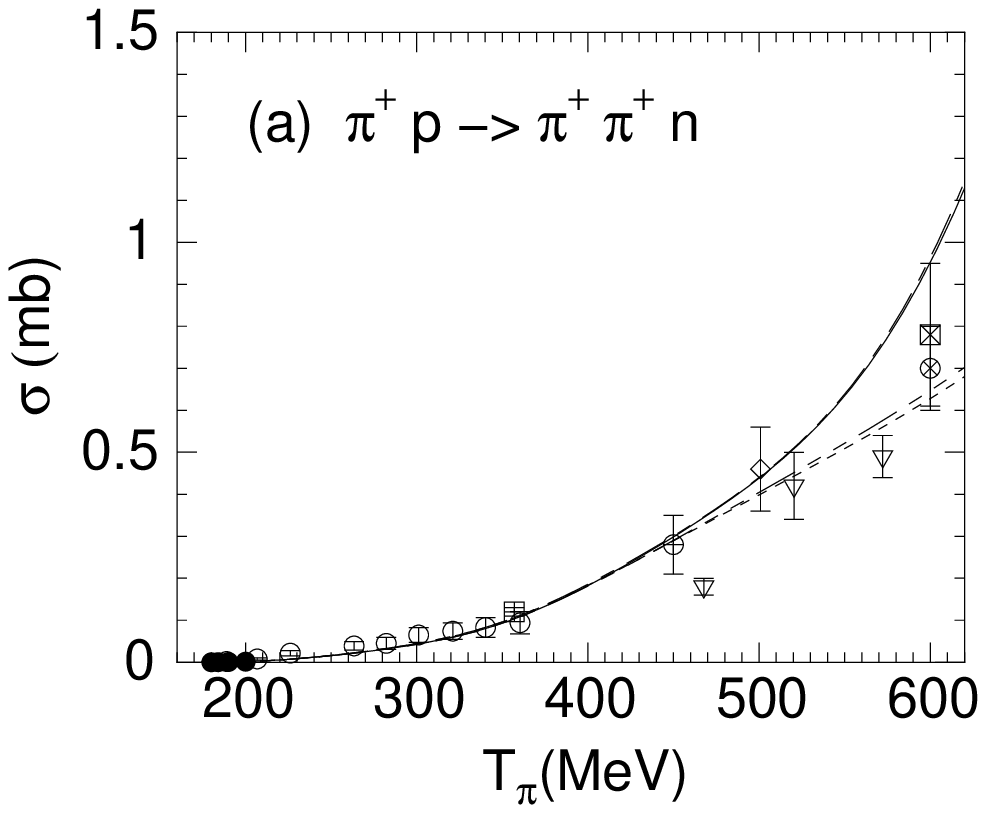}
\includegraphics[width=5cm]{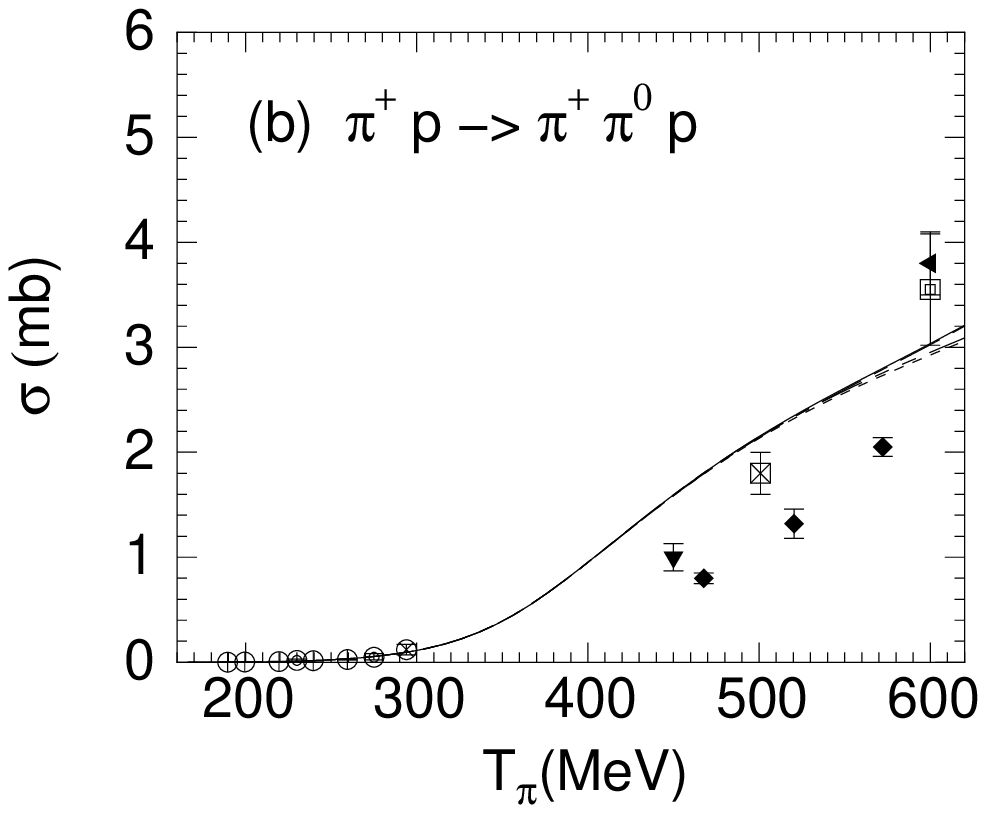}
\includegraphics[width=5cm]{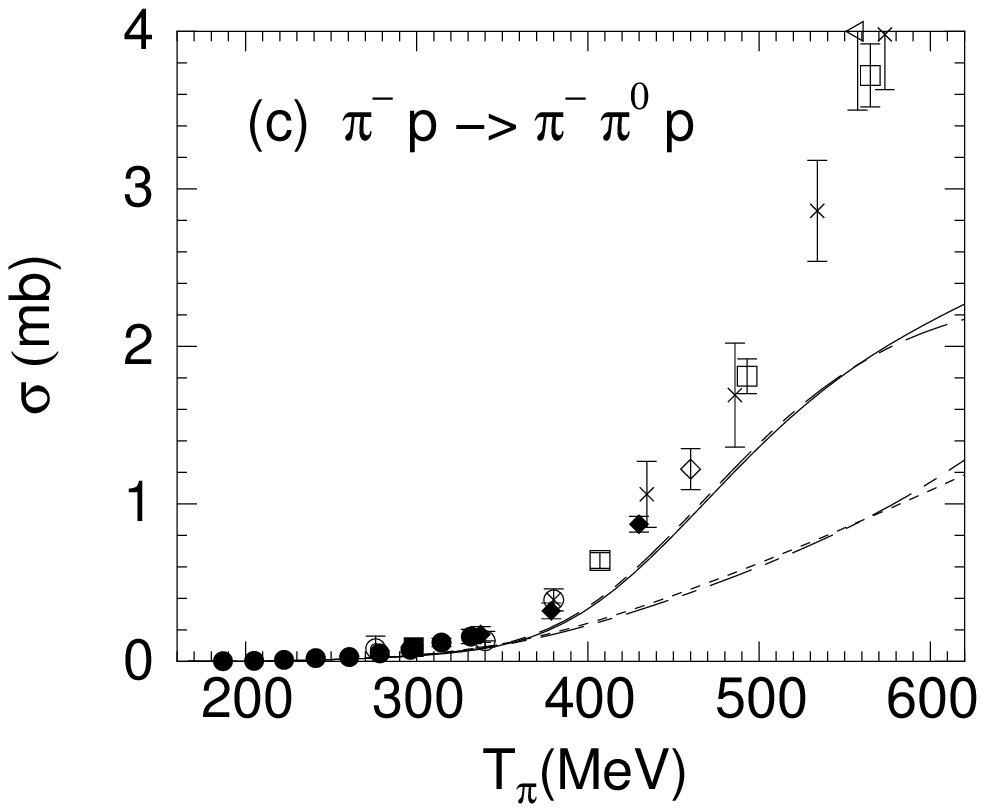}
\includegraphics[width=5cm]{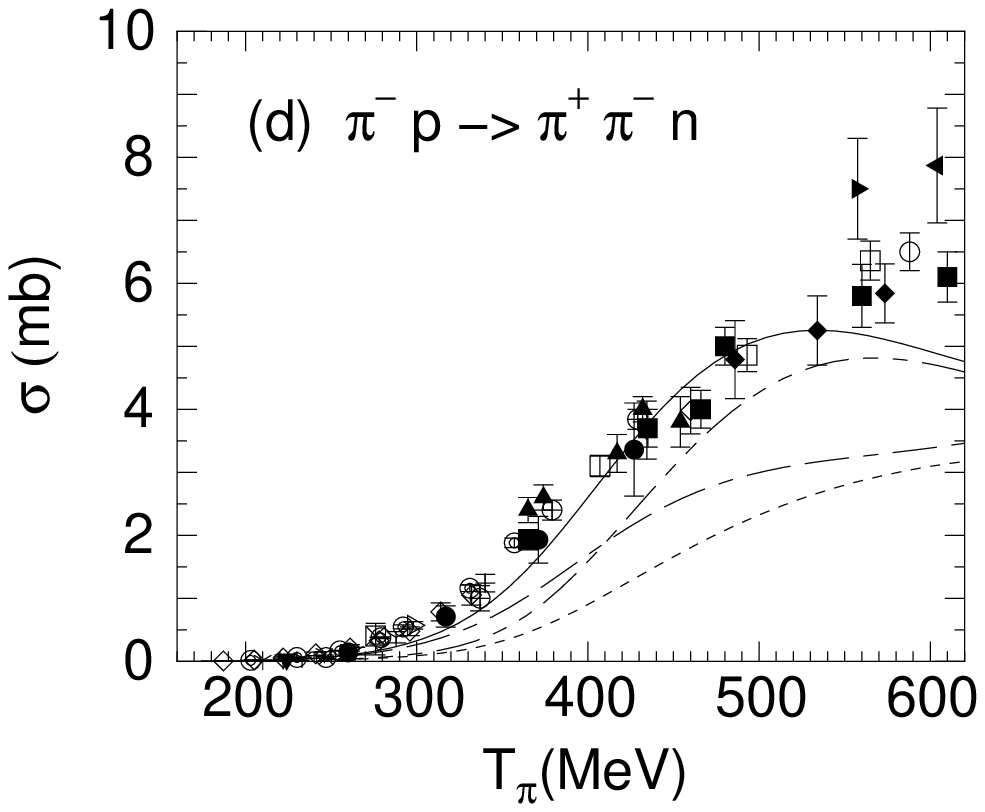}
\includegraphics[width=5cm]{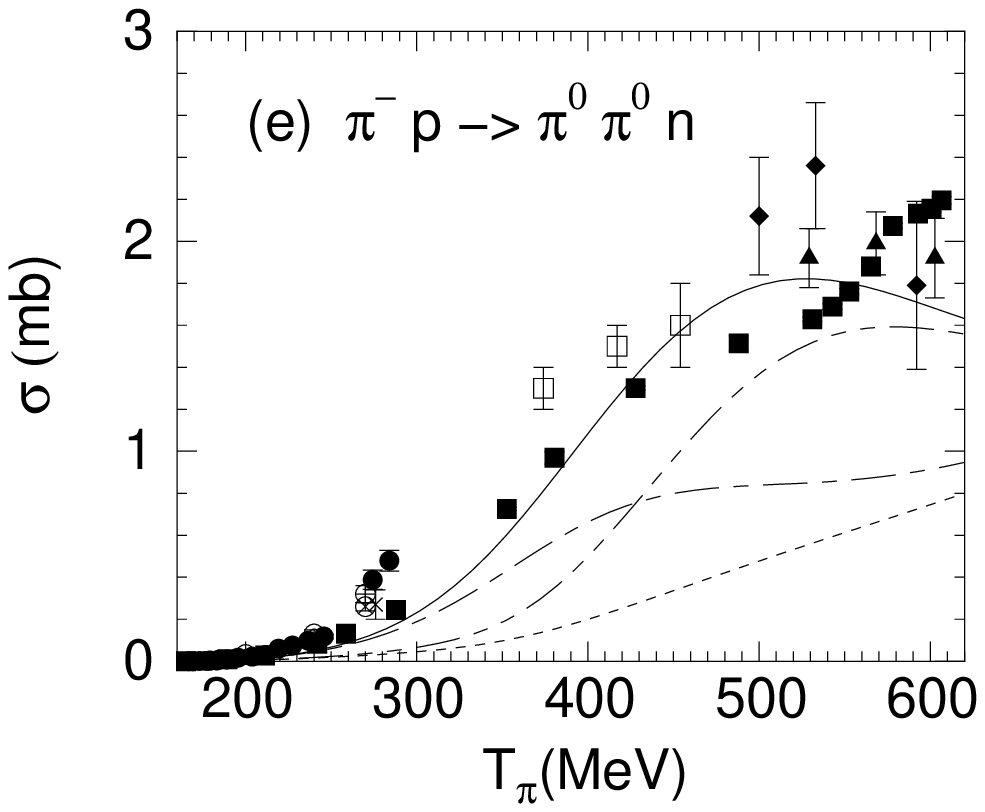}
\caption{
The contribution of each diagram including $N\sp{\ast}(1440)$ to the
total cross sections.
The value of ($c\sp{\ast}\sb{1},c\sp{\ast}\sb{2}$) is taken as 
$(-8.0,1.08)$~GeV$\sp{-1}$.
Each line expresses the result including 
(i) all diagrams (solid line), 
(ii) no $\pi \Delta N\sp{\ast}$ vertex (dotted-dashed line),
(iii) no $NN\sp{\ast}(\pi \pi)\sp{I=0}\sb{S\text{ wave}}$ vertex
(2-dotted-dashed line),
and 
(iv) neither contribution (dotted line).
The data are the same as in Fig.~\ref{fig9}.
}
\label{fig10}
\end{figure}
Now we discuss the contribution of 
each diagram including $N\sp{\ast}(1440)$.
The results are shown in Fig.~\ref{fig10}.
We find that the process Fig.~\ref{fig5}(b)
gives negligible contribution for all channels.
The process Fig.~\ref{fig5}(c) does not give visible influence on 
the $\pi\sp{+} p \rightarrow \pi\sp{+} \pi\sp{+} n$  
and $\pi\sp{+} p \rightarrow \pi\sp{+} \pi\sp{0} p$ channels,
but brings 30-50\% increase to the total cross section above 
$T\sb{\pi} = 400$~MeV for other three channels.

In the $\pi\sp{-} p \rightarrow \pi\sp{0} \pi\sp{0} n$ channel,
the process including
$NN\sp{\ast}(\pi\pi)\sp{I=0}\sb{S\text{ wave}}$ vertex dominates
the total cross sections near the threshold, 
whereas its contribution decreases gradually above $T\sb{\pi}=400$~MeV.
Instead the contribution of the process Fig.~\ref{fig5}(c)
grows in this energy region.
We find that the shoulder of 
the $\pi\sp{-} p \rightarrow \pi\sp{0} \pi\sp{0} n$ data 
at the $N\sp{\ast}(1440)$ energy reported in Ref.~\cite{Pra04}
can be understood as a result of interference between these two processes
rather than the sole contribution of $N\sp{\ast}(1440)$ pole.

With regard to the diagrams including $N\sp{\ast}(1440)$,
the $\pi\sp{-} p \rightarrow \pi\sp{+} \pi\sp{-} n$  
and $\pi\sp{-} p \rightarrow \pi\sp{0} \pi\sp{0} n$ channels
show similar behavior.
It is thus expected that 
the shoulder at $N\sp{\ast}(1440)$ energy may be observed 
also in the $\pi\sp{-} p \rightarrow \pi\sp{+} \pi\sp{-} n$ channel.
To see this, however, more accurate data in those energy region are necessary.

Here we notice that our result underestimates
the data above $T\sb{\pi}\sim 500$ MeV
in the $\pi\sp{-} p \rightarrow \pi\sp{-} \pi\sp{0} p$,  
$\pi\sp{-} p \rightarrow \pi\sp{+} \pi\sp{-} n$,  
and $\pi\sp{-} p \rightarrow \pi\sp{0} \pi\sp{0} n$ channels
(particularly remarkable for $\pi\sp{-} p \rightarrow \pi\sp{-} \pi\sp{0} p$). 
This is not surprising because that energy region is above the Roper
mass-shell energy. 
The higher resonances such as $N\sp{\ast}(1520)$ and $N\sp{\ast}(1535)$,
which do not consider in this work, will become relevant 
to these channels.
In particular, $N\sp{\ast}(1520)$ would play a key role
since this resonance has a large branching ratio about 45 \% for the
$\pi \pi N$ decay [whereas the $N\sp{\ast}(1535)$ has only 
about the 5~\% branching ratio].
Indeed, the total cross section data for these three reaction channels
show a small bump around the $N\sp{\ast}(1520)$ energy
(see e.g. Fig.~5(b) and (c) in Ref.~\cite{Man84} for
the $\pi\sp{-} p \rightarrow \pi\sp{-} \pi\sp{0} p$ and 
$\pi\sp{-} p \rightarrow \pi\sp{+} \pi\sp{-} n$ channels, 
and the CBC data~\cite{Pra04} for
$\pi\sp{-} p \rightarrow \pi\sp{0} \pi\sp{0} n$ channel.). 
However, such structure is not seen in the 
$\pi\sp{+} p \rightarrow \pi\sp{+} \pi\sp{+} n$ and 
$\pi\sp{+} p \rightarrow \pi\sp{+} \pi\sp{0} p$ channels
(see e.g. Fig.~5(d) and (e) in Ref.~\cite{Man84}). 
\subsection{Invariant mass distribution of 
$\pi\sp{-} p \rightarrow \pi\sp{0} \pi\sp{0} n$ reaction}
\label{sec3-3}
\begin{figure}
\includegraphics{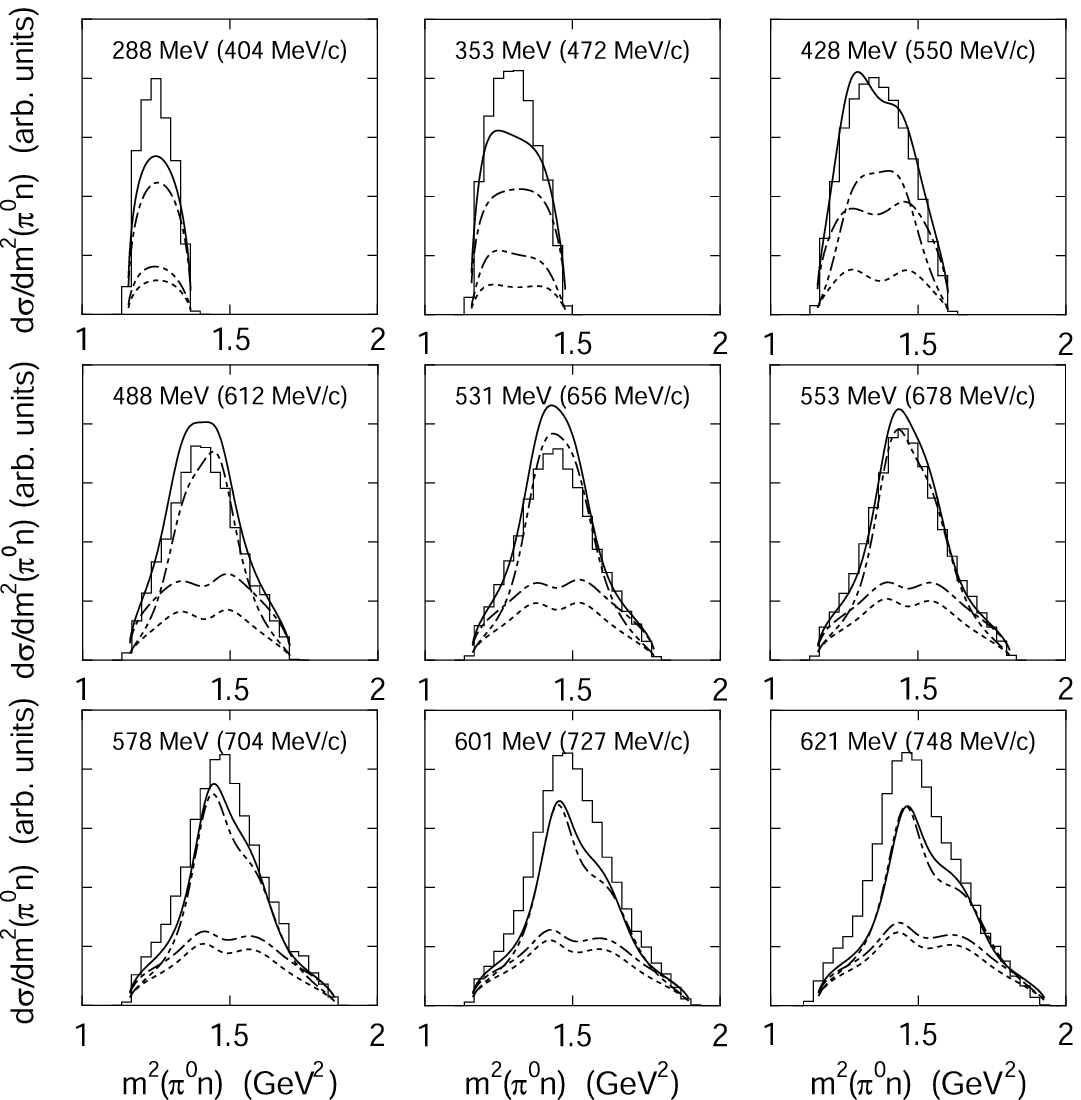}
\caption{
The $\pi\sp{0} n$ invariant mass distributions for 
the $\pi\sp{-} p \rightarrow \pi\sp{0} \pi\sp{0} n$ reaction
for several values of $T\sb{\pi}$ ($p\sb{\pi}$).
The CBC data~\cite{Pra04} are plotted as the histogram.
Each line expresses the results including 
(i) all diagrams (solid line),
(ii) no $\pi \Delta N\sp{\ast}$ vertex (dotted-dashed line),
(iii) no $NN\sp{\ast}(\pi \pi)\sp{I=0}\sb{S\text{ wave}}$ vertex
(2-dotted-dashed line),
and 
(iv) neither contribution (dotted line).
The $m\sp{2}(\pi\sp{0} n)= (1.210)\sp{2}=1.464$~GeV$\sp{2}$
and $m\sp{2}(\pi\sp{0} n)= (1.232)\sp{2}=1.518$~GeV$\sp{2}$ corresponds
to the real part of the $\Delta(1232)$ pole 
and the Breit-Wigner mass of $\Delta(1232)$,
respectively.
}
\label{fig11}
\end{figure}
Figure~\ref{fig11} shows the $\pi\sp{0} n$ invariant mass distributions
for several values of $T\sb{\pi}$. 
Our result captures the qualitative features of the data well.
At $T\sb{\pi}=288$~MeV, the mass distribution is almost determined by
the process including $NN\sp{\ast}(\pi\pi)\sp{I=0}\sb{S\text{ wave}}$ vertex. 
The contribution of this process decreases gradually 
when $T\sb{\pi}$ becomes higher.
Instead, the contribution of 
the process including $\pi\Delta N\sp{\ast}$ vertex 
dominates the mass distributions.
As suggested in Ref.~\cite{Pra04}, 
a peak near the $\Delta(1232)$ energy, 
which can be found above $T\sb{\pi}=488$~MeV,
is indeed generated by the $N\sp{\ast}(1440) \rightarrow \Delta \pi$ process.
Although the $\Delta(1232)$ mass is taken as $m\sb{\Delta} =1232$~MeV,
the peak occurs at the energy
$m\sp{2}(\pi\sp{0} n)\simeq (1.210)\sp{2}~\text{GeV}\sp{2}$.
In Ref.~\cite{Pra04} the authors said that the shift of the peak 
is natural because
the nucleon pole term, which is large background for
the $\pi N$ elastic channel, has small contribution to this reaction.
However, in view of our numerical result,
we find that the peak position is obtained
as a result of the interference of several processes.

\begin{figure}
\includegraphics{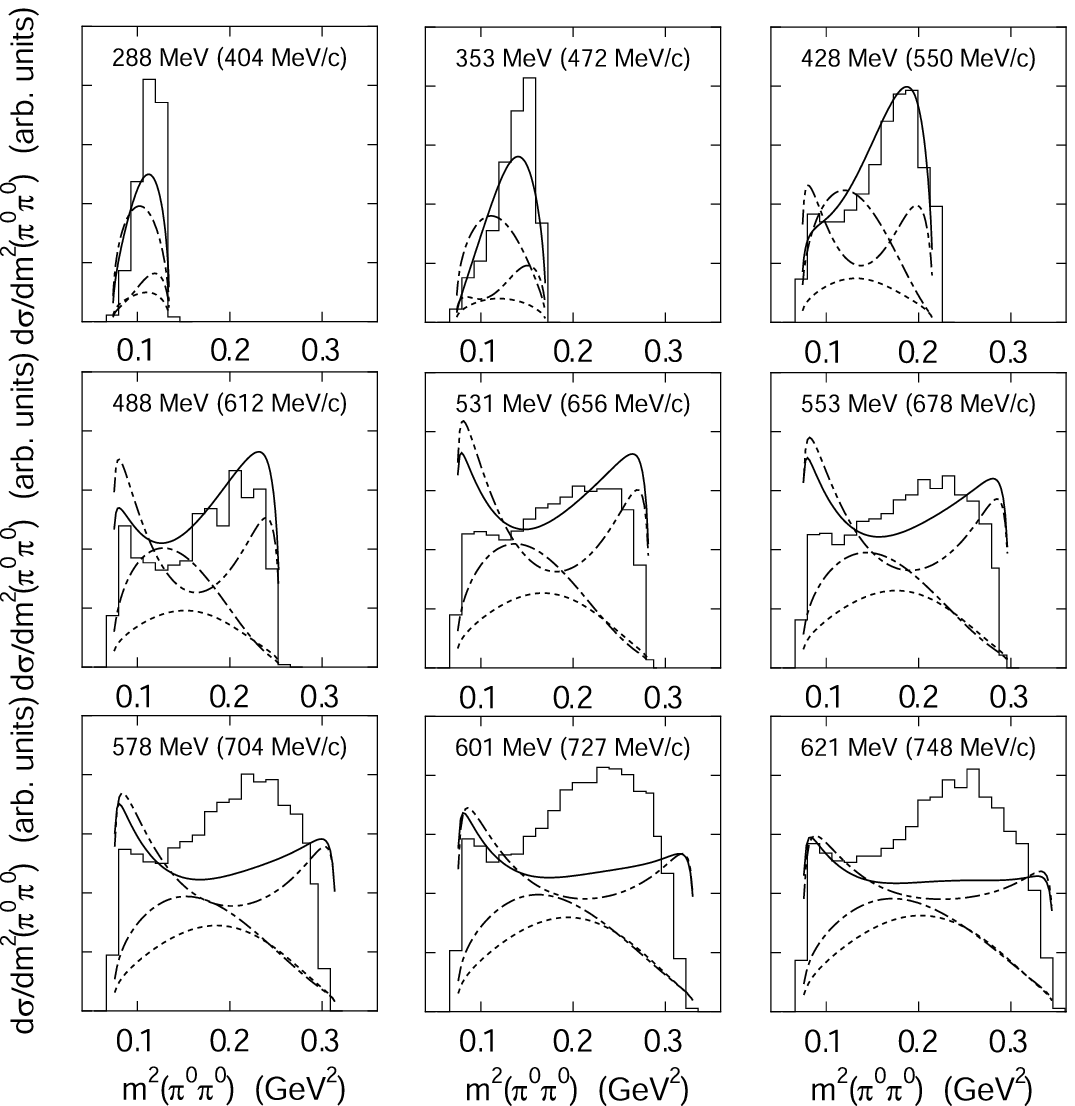}
\caption{
The $\pi\sp{0} \pi\sp{0}$ invariant mass distributions for 
the $\pi\sp{-} p \rightarrow \pi\sp{0} \pi\sp{0} n$ reaction
for several values of $T\sb{\pi}$ ($p\sb{\pi}$).
The meaning of each line and histogram is the same as in Fig.~\ref{fig11}.
}
\label{fig12}
\end{figure}
We next consider the $\pi\sp{0}\pi\sp{0}$ invariant mass distribution
(Fig.~\ref{fig12}).
For low $T\sb{\pi}$, the results are dominated by the process including
$NN\sp{\ast}(\pi\pi)\sb{S\text{ wave}}\sp{I=0}$ vertex,
which is the same as the case of $\pi\sp{0} n$ invariant mass distribution.
We observe that our result reproduces the large asymmetry 
in the mass distributions around the $N\sp{\ast}(1440)$ energy: 
a small (large) peak in small (large) value of $m\sp{2}(\pi\sp{0}\pi\sp{0})$
and a depletion in between 
(see the result of $T\sb{\pi}=488$~MeV in Fig.~\ref{fig12}).
In our results, the process with
$NN\sp{\ast}(\pi\pi)\sp{I=0}\sb{S\text{ wave}}$ vertex decreases 
the distribution for small value of $m\sp{2}(\pi\sp{0}\pi\sp{0})$,
whereas it increases for large value of $m\sp{2}(\pi\sp{0}\pi\sp{0})$.
Without this process, the large peak at large value of
$m\sp{2}(\pi\sp{0}\pi\sp{0})$ can not be reproduced.
This result indicates that the large asymmetry in the
$\pi\sp{0} \pi\sp{0}$ mass distribution at the $N\sp{\ast}(1440)$ energy
is due to the strong interference between 
the decay processes $N\sp{\ast}(1440)\rightarrow \Delta \pi$
and $N\sp{\ast}(1440)\rightarrow N(\pi\pi)\sb{S\text{ wave}}\sp{I=0}$.
Our numerical results up to $T\sb{\pi}=500$~MeV agree with the 
data qualitatively. 
Above $T\sb{\pi}=500$~MeV, however, our results 
do not reproduce a large peak at larger $m\sp{2}(\pi\sp{0}\pi\sp{0})$. 
This would be because the contributions of higher mass resonances
such as $N\sp{\ast}(1520)$ and $N\sp{\ast}(1535)$ are not taken into account.

It is worth mentioning that the above features of the 
$\pi\sp{0}\pi\sp{0}$ and $\pi\sp{0}n$ mass distributions 
at the $N\sp{\ast}(1440)$ energy are characteristic of
the two-pion decay of the Roper resonance, 
$N\sp{\ast}(1440) \rightarrow N \pi \pi$~\cite{Kam04,Her02}.
This indicates that $N\sp{\ast}(1440)$ indeed gives visible effect to 
the observables of $\pi N \rightarrow \pi \pi N$ reaction 
in those energy region.
\subsection{$I=J=0$ $\pi \pi$ rescattering
in $NN\sp{\ast}(\pi\pi)\sp{I=0}\sb{S\text{ wave}}$ vertex.
}
\label{sec3-4}
Finally we discuss the $\pi\pi$ rescattering representing
the scalar-isoscalar 
$\pi\pi$ correlation explicitly in
the $NN\sp{\ast}(\pi\pi)\sp{I=0}\sb{S\text{ wave}}$ vertex, 
which is depicted in Fig.~\ref{fig4}(b).
It has been suggested in several literatures 
(see e.g. Refs.~\cite{Kai98,Oll97}) that 
the $\sigma$ meson pole can be dynamically generated
around $450-225i$~MeV
by such rescattering in $I$=$J$=0 channel.
In view of the $\pi N \rightarrow \pi \pi N$ reaction 
being sensitive to the process including
$NN\sp{\ast}(\pi\pi)\sp{I=0}\sb{S\text{ wave}}$ vertex,
we can expect that this reaction also
becomes a source of information about this meson.
However, no direct evidence 
of $\sigma$ meson is seen in the measured data of 
$\pi\sp{-} p \rightarrow \pi\sp{0} \pi\sp{0} n$ reaction in Ref.~\cite{Pra04}.

\begin{figure}
\includegraphics[height=4cm]{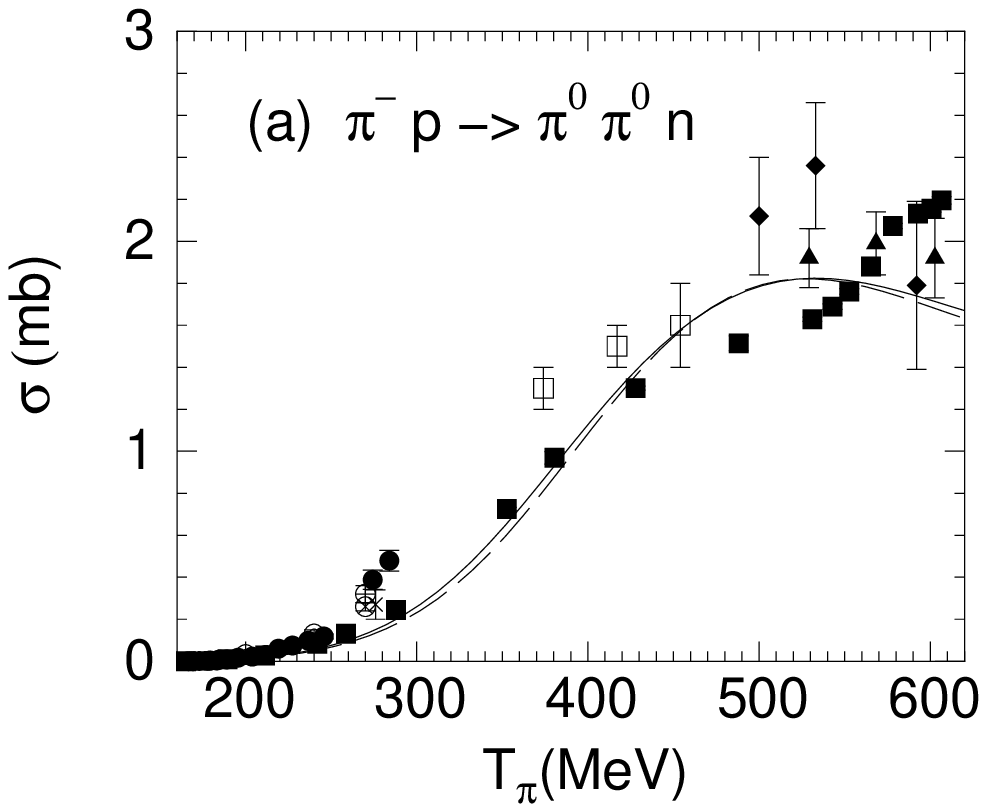}
\includegraphics[height=4cm]{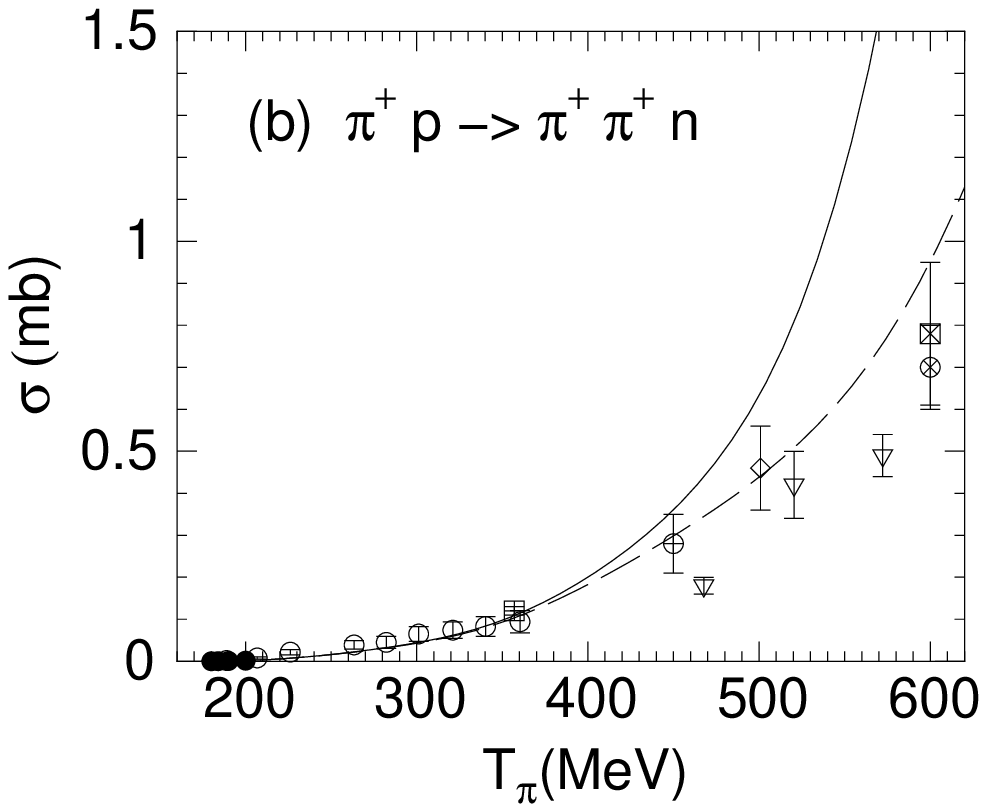}
\includegraphics[height=4cm]{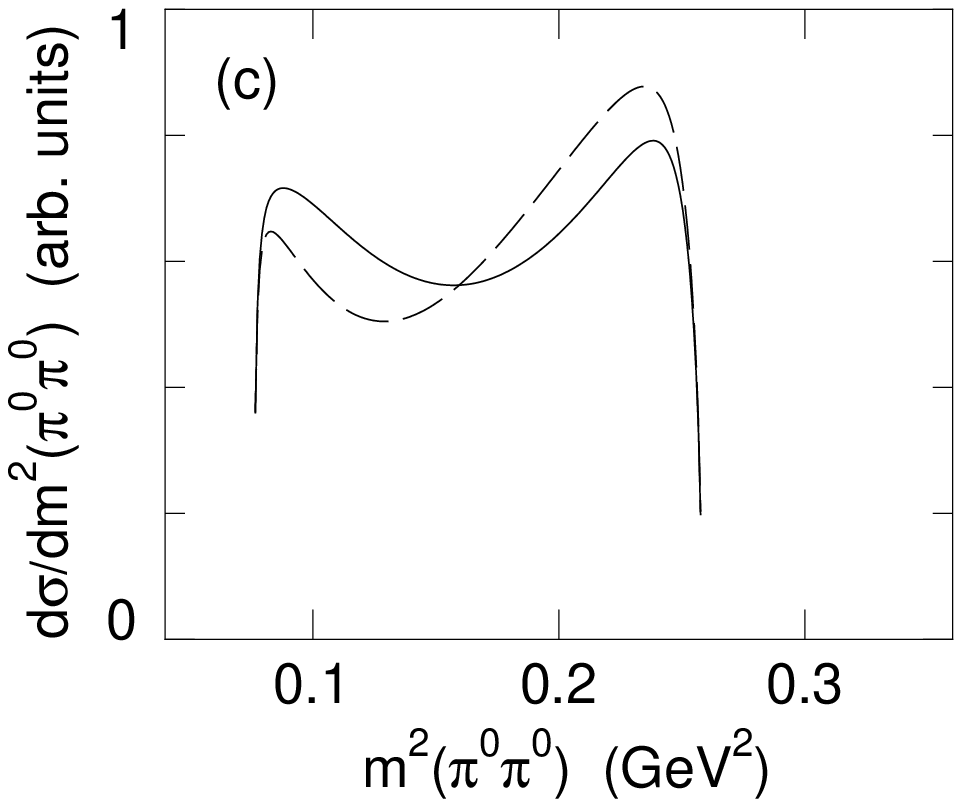}
\caption{
The $\pi \pi$ rescattering effect in the $\pi N \rightarrow \pi \pi N$ 
reaction:
the (a) $\pi\sp{-} p \rightarrow \pi\sp{0}\pi\sp{0} n$
and (b) $\pi\sp{+} p \rightarrow \pi\sp{+}\pi\sp{+} n$
total cross sections, and 
(c) the $\pi\sp{0}\pi\sp{0}$ invariant mass distribution
of $\pi\sp{-} p \rightarrow \pi\sp{0}\pi\sp{0} n$ at $T\sb{\pi}=488$~MeV.
The dashed line is the full result including the $\pi \pi$ rescattering
with $(c\sp{\ast}\sb{1},c\sp{\ast}\sb{2}) = (-8.0,1.08)$~GeV$\sp{-1}$,
whereas the solid line is the result including no $\pi\pi$ rescattering but
with the readjusted value 
$(c\sp{\ast}\sb{1},c\sp{\ast}\sb{2}) = (-13.0,1.89)$~GeV$\sp{-1}$.
}
\label{fig13}
\end{figure}

First we try to calculate
the $\pi\sp{-} p \rightarrow \pi\sp{0} \pi\sp{0} n$ total cross section
\textit{without} the process including $\pi\pi$ rescattering
in the $NN\sp{\ast}(\pi\pi)\sp{I=0}\sb{S\text{ wave}}$ vertex,
i.e. we consider only the contact interaction Fig.~\ref{fig4}(a)
for this vertex.
We find that, even if the $\pi\pi$ rescattering is not considered, 
we can readjust 
$c\sp{\ast}\sb{1}$ and $c\sp{\ast}\sb{2}$ 
within their allowed values so as to reproduce our results 
including the $\pi\pi$ rescattering effect\footnote{Note that in this case 
the coefficients in the decay width formula~(\ref{eq23}) become as 
$\alpha=0.476\times 10\sp{-3}$~GeV$\sp{3}$,
$\beta=5.29\times 10\sp{-3}$ ~GeV$\sp{3}$, 
and $\gamma=2.98\times 10\sp{-3}$~GeV$\sp{3}$. 
Thus the range of allowed values of $c\sp{\ast}\sb{1}$ and $c\sp{\ast}\sb{2}$ 
changes.}.
For instance, the total cross section with $c\sp{\ast}\sb{1}=-8.0$ 
[i.e. the solid line in Fig.~\ref{fig9}(e)] is reproduced
by choosing $c\sp{\ast}\sb{1}=-13.0$ and $c\sp{\ast}\sb{2}=1.89$
[see Fig.~\ref{fig13}(a)].
The similar results are also obtained for 
the $\pi\sp{-} p \rightarrow \pi\sp{+} \pi\sp{-} n$
and $\pi\sp{\pm} p \rightarrow \pi\sp{\pm} \pi\sp{0} p$ total cross sections,
and for the $\pi\sp{0}n$ invariant mass distributions of
$\pi\sp{-} p \rightarrow \pi\sp{0} \pi\sp{0} n$.
Therefore, although the $NN\sp{\ast}(\pi\pi)\sp{I=0}\sb{S\text{ wave}}$ vertex
is necessary to explain the $\pi N \rightarrow \pi \pi N$
reaction, it seems difficult to conclude
whether the $\sigma$ meson pole dominates the form factors of the vertices
as long as we consider only these total cross sections and invariant
mass distributions.

However, the situation is somewhat different in
the $\pi\sp{+} p \rightarrow \pi\sp{+} \pi\sp{+} n$ total cross section
and the $\pi\sp{0}\pi\sp{0}$ invariant mass distributions of
$\pi\sp{-} p \rightarrow \pi\sp{0} \pi\sp{0} n$.
The above value 
$(c\sp{\ast}\sb{1},c\sp{\ast}\sb{2})=(-13.0,1.89)$~GeV$\sp{-1}$ 
is not acceptable 
for the $\pi\sp{+} p \rightarrow \pi\sp{+} \pi\sp{+} n$ total cross section
because these values lead to the numerical results far from the data 
[Fig.~\ref{fig13}(b)].
Also, it is remarkable that the large asymmetric shape
in the $\pi\sp{0}\pi\sp{0}$ invariant mass distribution 
at $T\sb{\pi} = 488$~MeV can not be reproduced 
without considering the $\pi\pi$ rescattering effect.
These results suggest that the $\pi \pi$ rescattering effect 
in the $NN\sp{\ast}(\pi\pi)\sp{I=0}\sb{S\text{ wave}}$ vertex is 
\textit{necessary} to our calculation of $\pi N \rightarrow \pi\pi N$ reaction.
\section{Summary and Conclusions}
\label{sec4}
We have studied the $\pi N \rightarrow \pi \pi N$ reaction
in the energy region up to $T\sb{\pi}=620$~MeV, especially around 
the $N\sp{\ast}(1440)$ mass-shell energy. 
Being motivated by several interesting observations related to
$N\sp{\ast}(1440)$ in the recent CBC experiment
of the $\pi\sp{-}p\rightarrow \pi\sp{0}\pi\sp{0}n$ reaction,
we have discussed the role
of $N\sp{\ast}(1440)$ and its decay processes 
in the reaction processes.
The calculation has been performed by extending the theoretical approach
constructed in Ref.~\cite{Kam04}.

We have found that $N\sp{\ast}(1440)$ shows a significant contribution
to the $\pi N \rightarrow \pi \pi N$ reaction through the decay processes 
$N\sp{\ast}(1440) \rightarrow N(\pi\pi)\sp{I=0}\sb{S\text{ wave}}$
and $N\sp{\ast}(1440)\rightarrow \pi\Delta$. 
In contrast, the $N\sp{\ast}\rightarrow \pi N$ process 
just gives a negligible contribution.
While the contribution of 
$N\sp{\ast}(1440) \rightarrow N(\pi\pi)\sp{I=0}\sb{S\text{ wave}}$
process already appears in the threshold region in several channels, 
the $N\sp{\ast}(1440) \rightarrow \pi\Delta$ process
becomes important above $T\sb{\pi}=400$~MeV.
The characteristics of the CBC data 
for $\pi\sp{-} p \rightarrow \pi\sp{0} \pi\sp{0} n$ 
are generated by a strong interference effect between them.

We have also found that, above $T\sb{\pi}=400$~MeV,
the $\pi\sp{+} p \rightarrow \pi\sp{+} \pi\sp{+} n$
total cross section is remarkably sensitive to 
the variation of $c\sp{\ast}\sb{1}$ and $c\sp{\ast}\sb{2}$ 
within their allowed values.
Because this reaction is almost saturated by
the contributions of the nucleon and $\Delta(1232)$, 
more accurate data of this channel would give strong constraints 
on the range of $c\sp{\ast}\sb{1}$ and $c\sp{\ast}\sb{2}$.
This nature of $\pi\sp{+} p \rightarrow \pi\sp{+} \pi\sp{+} n$ 
can not be seen in other theoretical studies 
which have mainly focused on the $\pi N\rightarrow \pi \pi N$ reaction 
below $T\sb{\pi}=400$~MeV.

The remarkable contribution of the process including 
the $N\sp{\ast}N(\pi\pi)\sp{I=0}\sb{S\text{ wave}}$ vertex
leads to the expectation that
the $\pi N \rightarrow \pi \pi N$ reaction
becomes a source of information about
the controversial scalar-isoscalar $\sigma$ meson.
In the present work, 
the $\sigma$ meson is considered  as a dynamical object generated 
by the rescattering mechanism of two-pion in the $I=J=0$ channel.
We have obtained the following interesting results
related to this meson:
if we do not include the $\pi \pi$ rescattering effect in
the $N\sp{\ast}N(\pi\pi)\sp{I=0}\sb{\text{ wave}}$ vertex,
then (i) the large asymmetry in the 
$\pi\sp{0}\pi\sp{0}$ invariant mass distribution
for $\pi\sp{-} p \rightarrow \pi\sp{0} \pi\sp{0} n$
can not be reproduced, and 
(ii) it is difficult to simultaneously describe
the total cross section of all channels. 
The systematic analyses of all channels
would provide an indication of the existence of $\sigma$ meson.

Here we mention the similarity about 
the role of $N\sp{\ast}(1440)$
between the $\pi N \rightarrow \pi \pi N$ reaction and
the $NN$ induced two-pion production reactions.
The interference between
$N\sp{\ast}(1440) \rightarrow N(\pi\pi)\sp{I=0}\sb{S\text{ wave}}$
and $N\sp{\ast}(1440) \rightarrow \Delta\pi$
is observed also in the $pp \rightarrow pp \pi\sp{+}\pi\sp{-}$
and $pn \rightarrow d(\pi\pi)\sp{I=0}$ reactions~\cite{Alv99,Alv01},
and actually plays a important role for explaining the data~\cite{Bro02,Pat03}.
In view of this similarity, we could also discuss 
the indications of $\sigma$ meson through those reactions.

Finally, we comment on other baryon resonances related to
the $\pi N \rightarrow \pi \pi N$ reaction, 
which have been referred in several places.
In the present work, we did not include the higher resonances
such as $N\sp{\ast}(1520)$ and $N\sp{\ast}(1535)$.
We have seen that,
below the $N\sp{\ast}(1440)$ mass-shell energy (i.e. $T\sb{\pi}\alt 480$~MeV),
the $\pi N \rightarrow \pi \pi N$ data is almost saturated 
by the nucleon, $\Delta(1232)$ and $N\sp{\ast}(1440)$.
Therefore the higher resonances 
would become the relevant degrees of freedom
at least above $T\sb{\pi}\sim 480$~MeV.
Several discrepancies between the data and our results
above $T\sb{\pi}=500$~MeV
in the $\pi\sp{-} p \rightarrow \pi\sp{-} \pi\sp{0} p$, 
$\pi\sp{-} p \rightarrow \pi\sp{+} \pi\sp{-} n$ and
$\pi\sp{-} p \rightarrow \pi\sp{0} \pi\sp{0} n$ channels,
would be cured by the consideration of such higher resonances. 
Anyway, we need further investigations on these resonances.
\begin{acknowledgments}
The authors acknowledge the nuclear theory group of Osaka City University
for useful discussions.
The authors also would like to thank Dr.~Luis Alvarez-Ruso 
for valuable information.
H.~K. is supported by Research Fellowships of the Japan Society for
the Promotion of Science (JSPS) for Young Scientists.
\end{acknowledgments}
\appendix
\section{Details of the model in nucleon-$\Delta$ sector}
\label{app1}
As mentioned in Subsec.~\ref{sec2-1}, in this Appendix 
we explain some details of the model used in
our previous study of $\pi N \rightarrow \pi \pi N$~\cite{Kam04}.
To calculate the diagrams in Fig.~\ref{fig2},
we need to evaluate seven matrix elements shown in Table~\ref{tab1}.
\begin{table}
\caption{
The matrix elements necessary to calculate the diagrams in Fig.~\ref{fig2}.
}
\label{tab1}
\begin{ruledtabular}
\begin{tabular}{cccc}
$\bra{N}\vectorjc{}{}\ket{N}$ & $\bra{N}\axialjc{}{}\ket{N}$ &
$\bra{N}\hat{\sigma}\ket{N}$  & $\bra{N}\hat{\pi}\ket{N}$ \\
$\bra{\Delta}\vectorjc{}{}\ket{N}$ & $\bra{\Delta}\axialjc{}{}\ket{N}$ & & \\
& $\bra{\Delta}\axialjc{}{}\ket{\Delta}$ & & \\
\end{tabular}
\end{ruledtabular}
\end{table}
\subsection{Vector-isovector part}
\label{app1-1}
The vector current matrix elements are
\begin{equation}
\bra{N(p\sp{\prime})} \vectorjc{a}{\mu}(0) \ket{N(p)} =
\bar{u}(p\sp{\prime})
\left[ 
  F\sp{N}\sb{V,1}(t)\gamma\sb{\mu}
+ F\sp{N}\sb{V,2}(t)\frac{i}{2m\sb{N}}\sigma\sb{\mu\nu}q\sp{\nu}
\right]
 \frac{\tau\sp{a}}{2} u(p)
\label{eqa1}
\end{equation}
for the nucleon, and
\begin{eqnarray}
\bra{\Delta(p\sp{\prime})} \vectorjc{a}{\mu}(0) \ket{N(p)} &=&
\bar{U}\sp{\nu}(p\sp{\prime})
\left[
  F\sp{N\Delta}\sb{V,1}(t)g\sb{\nu\mu}
+ F\sp{N\Delta}\sb{V,2}(t)Q\sb{\nu}\gamma\sb{\mu}
\right.
\nonumber\\
&&
\left.
+  F\sp{N\Delta}\sb{V,3}(t)Q\sb{\nu}Q\sb{\mu}
+ iF\sp{N\Delta}\sb{V,4}(t)Q\sb{\nu}\sigma\sb{\mu\lambda}Q\sp{\lambda}
\right]
\gamma\sb{5} \isomtx{a}{3}{1} u(p)
\label{eqa2}
\end{eqnarray}
for the $N$-$\Delta$ transition.
In this appendix the isodoublet Dirac spinor for the nucleon is denoted as
$u(p)$.

Based on the phenomenology of vector meson dominance (VMD) 
that at low energy the matrix elements of vector current $\vectorjc{}{}$ 
are dominated by the $\rho$ meson pole,
we write the nucleon form factors as 
\begin{equation}
F\sp{N}\sb{V,1}(t) = 
\frac{m\sb{\rho}\sp{2}}
{m\sb{\rho}\sp{2}-t-im\sb{\rho}\Gamma\sb{\rho}(t)},\label{eqa4} 
\end{equation}
\begin{equation}
F\sp{N}\sb{V,2}(t) = \kappa\sb{V}F\sp{N}\sb{V,1}(t)
\label{eqa5},
\end{equation}
where $m\sb{\rho}$ is the $\rho$ meson mass and $\kappa\sb{V}$
is the isovector magnetic moment.
The phenomenological width of the $\rho$ meson
is parameterized as
\begin{equation}
\Gamma\sb{\rho}(t) = 
\Gamma\sb{\rho}\frac{m\sb{\rho}}{\sqrt{t}}
\left(
\frac{t - 4m\sb{\pi}\sp{2}}{m\sb{\rho}\sp{2} - 4m\sb{\pi}\sp{2}}\right)\sp{3/2}
\theta (t - 4m\sb{\pi}\sp{2}),
\label{eqa6}
\end{equation}
where $\Gamma\sb{\rho}$ is 
the total width at $t=m\sb{\rho}\sp{2}$~\cite{Ste97}.
As for the $N$-$\Delta$ transition form factors, we obtain
\begin{equation}
F\sp{N\Delta}\sb{V,1}(t) = 
\frac{f\sb{\rho N\Delta}}{f\sb{\rho}}
\left(
\frac{m\sb{N}+m\sb{\Delta}}{m\sb{\rho}}
\right)
\frac{m\sb{\rho}\sp{2}}{m\sb{\rho}\sp{2}-t-im\sb{\rho}\Gamma\sb{\rho}(t)},
\label{eqa7}
\end{equation}
\begin{equation}
F\sp{N\Delta}\sb{V,2}(t) = 
\frac{1}{m\sb{N}+m\sb{\Delta}} F\sp{N\Delta}\sb{V,1}(t),
\label{eqa8}
\end{equation}
where we use the $\rho N\Delta$ interaction (\ref{eqb5}).
The $\rho N\Delta$ coupling constant is denoted as $f\sb{\rho N\Delta}$,
and $f\sb{\rho}$ corresponds to the gauge coupling constant of 
the hidden local symmetry model for the vector mesons~\cite{Ban85,Har03}. 
The other form factors
$F\sp{N\Delta}\sb{V,3}(t)$ and $F\sp{N\Delta}\sb{V,4}(t)$ in Eq.~(\ref{eqa2})
are fixed to zero as long as we consider 
the Lagrangian (\ref{eqb5}) for the $\rho N\Delta$ interaction.
\subsection{Axial-isovector part}
The matrix elements of $\axialjc{}{}$ are written as~\cite{Kac96, Arn79}
\begin{equation}
\bra{N(p\sp{\prime})}\axialjc{a}{\mu}(0)\ket{N(p)} =
\bar{u}(p\sp{\prime})
\left[
  F\sp{N}\sb{A,1}(t)\gamma\sb{\mu}
+ F\sp{N}\sb{A,2}(t)q\sb{\mu}
\right]\gamma\sb{5}\frac{\tau\sp{a}}{2} u(p),
\label{eqa9}
\end{equation}
\begin{eqnarray}
\bra{\Delta(p\sp{\prime})}\axialjc{a}{\mu}(0)\ket{N(p)} &=&
\bar{U}\sp{\nu}(p\sp{\prime})
\left[
  F\sp{N\Delta}\sb{A,1}(t)g\sb{\nu\mu}
+ F\sp{N\Delta}\sb{A,2}(t)Q\sb{\nu}\gamma\sb{\mu}
\right.
\nonumber\\
&&
\left.
+  F\sp{N\Delta}\sb{A,3}(t)Q\sb{\nu}Q\sb{\mu}
+ iF\sp{N\Delta}\sb{A,4}(t)Q\sb{\nu}\sigma\sb{\mu\lambda}Q\sp{\lambda}
\right] \isomtx{a}{3}{1} u(p),
\label{eqa10}
\end{eqnarray}
\begin{eqnarray}
\bra{\Delta(p\sp{\prime})}\axialjc{a}{\mu}(0)\ket{\Delta(p)} &=&
\bar{U}\sp{\nu}(p\sp{\prime})
\left[
  F\sp{\Delta}\sb{A,1}(t)g\sb{\nu\lambda}\gamma\sb{\mu}
+ F\sp{\Delta}\sb{A,2}(t)g\sb{\nu\lambda}q\sp{\mu}
\right.
\nonumber\\
&&
\left.
+ F\sp{\Delta}\sb{A,3}(t)(q\sb{\nu}g\sb{\mu\lambda}
+ g\sb{\nu\mu}q\sb{\lambda})
\right.
\nonumber\\
&&
\left.
+ F\sp{\Delta}\sb{A,4}(t)q\sb{\nu}\gamma\sb{\mu}q\sb{\lambda}
+ F\sp{\Delta}\sb{A,5}(t)q\sb{\nu}q\sb{\mu}q\sb{\lambda}
\right] \gamma\sb{5}\isomtx{a}{3}{3} U\sp{\lambda}(p).
\label{eqa11}
\end{eqnarray}

Same as in the case of $N$-$N\sp{\ast}$ and $\Delta$-$N\sp{\ast}$ transitions
caused by the axial current $\axialjc{}{}$,
some of the form factors in Eqs.~(\ref{eqa9})-(\ref{eqa11})
are exactly related to the renormalized coupling constants for 
the corresponding pion-baryon interaction,
\begin{equation}
f\sb{\pi N N}(t) =
\frac{m\sb{\pi}}{f\sb{\pi}}
\left[
  \frac{1}{2}F\sp{N}\sb{A,1}(t)
+ \frac{t}{4m\sb{N}}F\sp{N}\sb{A,2}(t)
\right],
\label{eqa12}
\end{equation}
\begin{equation}
f\sb{\pi N\Delta}(t) =
\frac{m\sb{\pi}}{f\sb{\pi}}
\left[
  F\sp{N\Delta}\sb{A,1}(t)
+ (m\sb{N}-m\sb{\Delta})F\sp{N\Delta}\sb{A,2}(t)
+ tF\sp{N\Delta}\sb{A,3}(t)\right],
\label{eqa13}
\end{equation}
\begin{equation}
f\sb{\pi\Delta\Delta}(t) =
\frac{m\sb{\pi}}{f\sb{\pi}}
\left[
  F\sp{\Delta}\sb{A,1}(t)
+ \frac{t}{2m\sb{\Delta}}F\sp{\Delta}\sb{A,2}(t)
\right],
\label{eqa14}
\end{equation}
where we employ Eqs.~(\ref{eqb1})-(\ref{eqb3}) as the effective 
$\pi NN$, $\pi N\Delta$ and $\pi\Delta\Delta$ interactions.
The other form factors  
(i.e. $F\sp{N\Delta}\sb{A,4}(t)$, $F\sp{\Delta}\sb{A,3}(t)$, 
$F\sp{\Delta}\sb{A,4}(t)$ and $F\sp{\Delta}\sb{A,5}(t)$)
are fixed to zero.
In our calculation all the form factors are taken as constants,
and we eliminate 
$F\sp{N\Delta}\sb{A,3}$ and $F\sp{\Delta}\sb{A,2}$ 
by using the PCAC hypothesis 
$f\sb{\pi N\Delta, \pi\Delta\Delta}(m\sb{\pi}\sp{2})\simeq 
f\sb{\pi N\Delta, \pi\Delta\Delta}(0)$.
\subsection{Scalar-isoscalar part}
Since $\Delta(1232)$ dose not contribute to ${\cal M}\sb{SA}$, 
we only need the nucleon matrix element for the $\hat{\sigma}$ current,
\begin{equation}
\bra{N(p\sp{\prime})} \hat{\sigma}(0) \ket{N(p)} =
S(t)\overline{u}(p\sp{\prime})u(p).
\label{eqa15}
\end{equation}
According to the definition in Ref.~\cite{Yam96}, 
the form factor $S(t)$ is equal to 
$-\sigma\sb{\pi N}(t)/ f\sb{\pi}m\sb{\pi}\sp{2}$, 
where $\sigma\sb{\pi N}(t)$ is the pion-nucleon sigma term 
which becomes independent of $t$ at tree level. 
\subsection{Pseudoscalar-isovector part}
The nucleon matrix element of the $\hat{\pi}$ current appearing
in ${\cal M}\sb{\pi}$ is written as
\begin{equation}
\bra{N(p\sp{\prime})}\hat{\pi}\sp{a}(0)\ket{N(p)} =
P(t)\overline{u}(p\sp{\prime})i\gamma\sb{5}\tau\sp{a}u(p).
\label{eqa16}
\end{equation}
The form factor $P(t)$ are related  to 
the renormalized $\pi NN$ coupling constant (\ref{eqa11}),
\begin{eqnarray}
P(t) =
\frac{1}{\mpi\sp{2} - t}
\left( \frac{2m\sb{N}}{m\sb{\pi}} \right)f\sb{\pi NN}(t),
\label{eq:28}
\end{eqnarray}
where the pion pole contribution is taken into account. 

In Table.~\ref{tab2}, we summarize the constants necessary to calculate 
the diagrams in Fig.~\ref{fig2}.
\begin{table}
\caption{
The value of constants used in the previous work.  
The mass and width of each particle or resonance are shown in MeV.
}
\label{tab2}
\begin{ruledtabular}
\begin{tabular}{cccc}
Masses and widths       &(MeV)  &
Parameters              &         \\ \hline
$m\sb{N}$               &939    &
$f\sb{\pi}$             &93~MeV   \\
$m\sb{\pi}$             &138    &
$f\sb{\rho}$            &5.80\footnotemark[1]\\
$m\sb{\Delta}$          &1232   &
$\kappa\sb{V}$          &3.71\\
$m\sb{\rho}$            &770    &
$\sigma\sb{\pi N}$      &45~MeV\footnotemark[2]\\
$\Gamma\sb{\Delta}$     &120    &
$F\sb{A,1}\sp{N}~(=g\sb{A})$               &1.265    \\
$\Gamma\sb{\rho}$       &149    &
$F\sb{A,2}\sp{N}$  &$5.67\times 10\sp{-3}$~MeV$\sp{-1}$\footnotemark[3]\\
                      &       &
$F\sp{N\Delta}\sb{A,1}$ &1.382\footnotemark[4]\\
                        &       &
$F\sp{N\Delta}\sb{A,2}$ &\ $-4.24\times 10\sp{-4}$ MeV$\sp{-1}$\footnotemark[4]
\end{tabular}
\end{ruledtabular}
\footnotetext[1]{See p.33 in Reference~\cite{Har03}}
\footnotetext[2]{Reference~\cite{Gas91}}
\footnotetext[3]{Reference~\cite{Ste98}. Note that the relation
$F\sb{A,2}\sp{N}=-2\overline{\Delta}\sb{\pi N}/\mpi\sp{2}$. }
\footnotetext[4]{Reference~\cite{Kac96}}
\end{table}
\section{Phenomenology of meson-baryon system}
\label{app2}
In this Appendix, we summarize the phenomenological Lagrangians 
used to estimate the form factors and mention a treatment of the finite width
of baryon resonances.
The Lagrangians of meson-baryon interaction are written as follows,
\begin{equation}
{\cal L}\sb{\pi NN} =
\frac{f\sb{\pi N N}}{\mpi}
\bar{N} \gamma\sb{\mu}\gamma\sb{5} \tau\sp{a} N
\partial\sp{\mu} \pi\sp{a},
\label{eqb1}
\end{equation}
\begin{equation}
{\cal L}\sb{\pi N\Delta} =
\frac{f\sb{\pi N \Delta}}{\mpi}
\bar{\Delta}\sp{\nu} \Theta\sb{\nu\mu}(Z\sb{1}) \isomtx{a}{3}{1} N
\partial\sp{\mu} \pi\sp{a} 
+ \text{H.c.} \ ,
\label{eqb2}
\end{equation}
\begin{equation}
{\cal L}\sb{\pi\Delta\Delta} =
\frac{f\sb{\pi \Delta \Delta}}{\mpi}
\bar{\Delta}\sp{\alpha} 
\Theta\sb{\alpha\beta}(Z\sb{2})
\gamma\sb{\mu}\gamma\sb{5} \isomtx{a}{3}{3} 
{\Theta\sp{\beta}}\sb{\delta}(Z\sb{2})
\Delta\sp{\delta} 
\partial\sp{\mu}\pi\sp{a},
\label{eqb3}
\end{equation}
\begin{equation}
{\cal L}\sb{\rho N \Delta} =
i\frac{f\sb{\rho N \Delta}}{m\sb{\rho}}
\bar{\Delta}\sp{\sigma}
\Theta\sb{\sigma\mu}(Z\sb{3})
\gamma\sb{\nu}\gamma\sb{5} \isomtx{a}{3}{1} 
N
(\partial\sp{\nu}\rho\sp{\mu a}-\partial\sp{\mu}\rho\sp{\nu a})
+ \text{H.c.} \ ,
\label{eqb5}
\end{equation}
\begin{equation}
{\cal L}\sb{\pi N N\sp{\ast}} =
\frac{f\sb{\pi NN\sp{\ast}}}{\mpi}
\bar{N}\gamma\sb{\mu}\gamma\sb{5}
\tau\sp{a}N\sp{\ast}\partial\sp{\mu}\pi\sp{a}
+ \text{H.c.},
\label{eqb6}
\end{equation}
\begin{equation}
{\cal L}\sb{\pi \Delta N\sp{\ast}} =
\frac{f\sb{\pi \Delta N\sp{\ast}}}{\mpi}
\bar{\Delta}\sp{\nu}
\Theta\sb{\nu\mu}(Z\sb{4})
I\sp{a}({\textstyle \frac{3}{2},\frac{1}{2}})
N\sp{\ast}\partial\sp{\mu}\pi\sp{a}
+ \text{H.c.},
\label{eqb7}
\end{equation}
where $N$ and $N\sp{\ast}$ are the isodoublet Dirac fields describing
the nucleon and $N\sp{\ast}(1440)$, respectively, and $\Delta\sp{\mu}$
is the isoquadruplet Rarita-Schwinger field describing the $\Delta(1232)$.
The second-rank Lorentz tensor $\Theta\sb{\mu\nu}$ is defined by 
$\Theta\sb{\mu\nu}(Z)\equiv g\sb{\mu\nu}-
\frac{1}{2}(1+2Z)\gamma\sb{\mu}\gamma\sb{\nu}$ 
(where we take $A=-1$ \cite{Nat71}). 
The second term of this tensor vanishes if $\Delta(1232)$ is on the mass-shell
(because of $\gamma\sb{\mu}U\sp{\mu}(p)=0$), 
and so $Z$ is called the off-shell parameter. 
In this paper we assume $Z\sb{i}=-1/2\ (i=1...4)$ for simplicity, i.e.
$\Theta\sb{\mu\nu}\rightarrow g\sb{\mu\nu}$.

The $\Delta(1232)$ propagator is 
\begin{equation}
S\sb{\mu\nu}(p) =
\frac{(\not{p} + m\sb{\Delta})}{3(p\sp{2} - m\sb{\Delta}\sp{2})}
\left[
- 2g\sb{\mu\nu} 
+ \frac{2p\sb{\mu}p\sb{\nu}}{m\sb{\Delta}\sp{2}}
- i\sigma\sb{\mu\nu}
+ \frac{\gamma\sb{\mu}p\sb{\nu} - \gamma\sb{\nu}p\sb{\mu}}{m\sb{\Delta}}
\right].
\label{eqb8}
\end{equation}
The nucleon and $N\sp{\ast}(1440)$ propagators are 
expressed by the Dirac propagator which is familiar.

In order to take account of the width of baryon resonances phenomenologically,
we modify the denominator of the $\Delta(1232)$ and $N\sp{\ast}(1440)$ 
propagators as
$p\sp{2} - m\sb{R}\sp{2} \rightarrow 
 p\sp{2} - m\sb{R}\sp{2} + im\sb{R}\Gamma\sb{R}(s)$,
where $R = \Delta$ or $N\sp{\ast}$.
The widths $\Gamma\sb{\Delta}(s)$ and 
$\Gamma\sb{N\sp{\ast}}(s)$ are taken as~\cite{Jen97,Ose85}
\begin{equation}
\Gamma\sb{\Delta}(s)=\Gamma\sb{\Delta}
\frac{m\sb{\Delta}}{\sqrt{s}}
\frac{|{\bf q}(\sqrt{s})|\sp{3}}{|{\bf q}(m\sb{\Delta})|\sp{3}}
\theta(\sqrt{s}-m\sb{N}-\mpi),
\label{eqb9}
\end{equation}
and
\begin{equation}
\Gamma\sb{N\sp{\ast}}(s)
= \Gamma\sb{N\sp{\ast}}
\frac{|\bm{q}(\sqrt{s})|\sp{3}}{|\bm{q}(m\sb{N\sp{\ast}})|\sp{3}}
\theta(\sqrt{s}-m\sb{N}-\mpi),
\label{eqb10}
\end{equation}
respectively. 
Here $\bm{q}=\bm{q}(\sqrt{s})$ is the pion spatial momentum in the center of
mass $\pi N$ system with the total energy $\sqrt{s}$.
In this paper we use $m\sb{N\sp{\ast}}=1440$~MeV and 
$\Gamma\sb{N\sp{\ast}}=350$~MeV 
(as for the value of 
$m\sb{\Delta}$ and $\Gamma\sb{\Delta}$, see Table.~\ref{tab2}).

Finally we list the formulae for the isospin matrices 
used in our calculation
(see e.g. appendix A in Ref. \cite{Ose85}),
\begin{equation}
\isomtx{a\dag}{3}{1}\isomtx{b}{3}{1} = 
\delta\sp{ab} - \frac{1}{3}\tau\sp{a}\tau\sp{b},
\label{eqb11}
\end{equation}
\begin{equation}
\isomtx{a\dag}{3}{1}\isomtx{b}{3}{3}\isomtx{c}{3}{1} = 
  \frac{5}{6}i\varepsilon\sp{abc}
- \frac{1}{6}\delta\sp{ab}\tau\sp{c}
+ \frac{2}{3}\delta\sp{ac}\tau\sp{b}
- \frac{1}{6}\delta\sp{bc}\tau\sp{a}.
\label{eqb12}
\end{equation}
%
%
%
%
%

%
%
%
%
\end{document}